\documentclass[prx,twocolumn,superscriptaddress,amsmath]{revtex4-2} 
\usepackage{mathrsfs}
\usepackage{amsfonts}
\usepackage{amsmath}
\usepackage{appendix}
\usepackage{txfonts}
\usepackage{amssymb}
\usepackage{graphicx,subfigure}
\usepackage{bm}
\usepackage{color}
\usepackage[normalem]{ulem}
\usepackage{dcolumn} 
\usepackage[table]{xcolor}
\usepackage{float}
\usepackage{tabularx}
\usepackage{hyperref}
\usepackage{bbm}
\newcommand{\ket}[1]{|#1\rangle}
\newcommand{\bra}[1]{\langle #1|}

\newcommand*{\QEDB}{\null\nobreak\hfill\ensuremath{\square}}

\begin{document}

\title{Nonlinear quantum multi-spin dynamics without entanglement} 

\author{Ivan P. Miranda}
\affiliation{Department of Mathematics and Physics, Linnaeus University, SE-39231 Kalmar, Sweden}
\email{ivan.depaulamiranda@lnu.se}

\author{Erik Sj\"oqvist}
\affiliation{Department of Physics and Astronomy, Uppsala University, 
Box 524, SE-751 20 Uppsala, Sweden}
\email{erik.sjoqvist@physics.uu.se}

\date{\today}

\begin{abstract}
We introduce a distinct route for extending classical atomistic spin dynamics into the quantum regime for arbitrary spin $s$, based on a variational product-state \textit{ansatz} that includes both pure and mixed local states. By construction, intersite entanglement -- expected to remain negligible even on very short time scales in many real materials -- is excluded, while local quantum effects are retained through the nonclassical structure of $s\geq 1$ density operators. Nonlinear dynamics arise from both the mean-field interdependence of the local states and the state-dependent dissipative terms. The framework recovers classical spin dynamics for 
$s=\frac{1}{2}$ while allowing genuinely quantum local dynamics for higher spins. It clarifies the variational and geometric origin of the inequivalence between quantum Landau-Lifshitz and Landau-Lifshitz-Gilbert dynamics, provides a criterion for their equivalence under a common time rescaling, establishes connections with the classical theory through two qualitatively distinct limits ($s\to\infty$ and a large number of quantum spins $N$), and allows realistic $\mathcal{O}(N)$ simulations of quantum spin dynamics in materials.
\end{abstract}

\maketitle

\section{Introduction}

The Landau-Lifshitz (LL) \cite{landau35} and Landau-Lifshitz-Gilbert (LLG) \cite{gilbert04} equations form the foundation of continuum and atomistic descriptions of magnetization dynamics in magnetic materials \cite{antropov96,eriksson17,Abert2019}. Their success rests on a remarkably economical description: each atomic magnetic moment is represented by a three-component classical vector whose dynamics results from precession in an effective field and a dissipative alignment toward that field. The mutual dependence of these effective fields on all other spins generates the nonlinear collective dynamics characteristic of magnetic materials, while the cross-product structure of both the conservative and dissipative terms preserves the magnitude of every classical spin.

This description, however, has a fundamental limitation. An atomic magnetic moment is a quantum object, and replacing it by a three-component classical vector discards local quantum degrees of freedom that can remain dynamically relevant even when quantum correlations between different sites are weak. The distinction becomes unavoidable for spin $s\geq 1$. A local density operator describing such systems contains degrees of freedom, including higher-rank multipolar fluctuations and coherences \cite{bharath19,zimba06,Baguette2017}, that cannot be represented by the expectation values of the three spin components alone. Thus, quantum corrections to atomistic spin dynamics need not originate from intersite entanglement: they can arise intrinsically from the geometry and dimensionality of the local quantum state space.

A fully quantum treatment in principle captures these effects. The state of an $N$-spin system is described by a density operator $\varrho(t)$ containing both local quantum coherences and intersite correlations \cite{wieser13,liu24,Ataei2026,Ataei2026b,Ataei2026c}. The price is the exponential growth of the Hilbert space dimension with system size. Even methods for evolving low-rank density operators are consequently restricted to small systems \cite{mirzaei26}. This exponential complexity creates a gap between two widely used descriptions: classical atomistic spin dynamics, which is scalable but contains no local quantum structure, and fully quantum many-body dynamics, which is microscopically complete but rapidly becomes computationally obstructive as the system grows.

To address this latter point, we will here take a different approach to finding the quantum analog dynamics of the classical LL and LLG. We ask what quantum spin dynamics remains when intersite entanglement is deliberately excluded while the local quantum state is retained in full. To this end, we use the variational product-state {\it ansatz} $\varrho(t) = \bigotimes_{j=1}^{N} \varrho_j(t)$, where each $\varrho_j$ is an arbitrary pure or mixed density operator on the local spin-$s$ Hilbert space. This defines a dynamical manifold that is exponentially smaller than the full many-body state space, yet is substantially richer than the classical phase space of $3N$ spin components. In particular, the local states retain the nonclassical structure of the full $(2s+1)$-dimensional Hilbert space, while the intersite correlations are restricted to those generated self-consistently through the mean fields produced by the local states. This construction is motivated both physically and conceptually. In realistic materials, localized spins interact with a large number of environmental and electronic degrees of freedom. Such coupling provides efficient channels for decoherence and can strongly suppress intersite quantum correlations on time scales relevant to magnetization dynamics. 

The product-state manifold provides a natural setting in which to address the relationship between LL and LLG dynamics. In classical spin dynamics, the LL and LLG equations are equivalent formulations related by a simple rescaling of time \cite{lakshmanan84}. This equivalence is not fundamental: it relies on the geometry of classical spin phase space and on the fixed magnitude of each spin. Once the local quantum state is allowed to be mixed, the variational structures underlying the two equations become different. We derive the quantum Landau-Lifshitz-Gilbert ($q$-LLG) equation using the Dirac-Frenkel variational principle supplemented by a Rayleigh-Lagrange-d'Alembert construction, while the corresponding quantum Landau-Lifshitz ($q$-LL) equation follows from an Onsager gradient-flow formulation. The resulting equations are, in general, inequivalent. We identify the geometric origin of this inequivalence and derive a criterion specifying when the two dynamics nevertheless generate the same trajectories after a common rescaling of time. Thus, the distinction between $q$-LL and $q$-LLG reflects the different ways in which reversible and dissipative dynamics are constructed on the quantum state manifold.

An important question is whether such a quantum extension retains a meaningful connection to the classical theory. We show that it does, but in two qualitatively different ways. The first is the conventional semiclassical route, obtained by taking the local spin quantum number $s$ to infinity. In this limit, the relative importance of the additional local quantum degrees of freedom diminishes and classical spin dynamics is recovered. The second route is fundamentally different: the number of quantum spins $N$ is increased while the local spin remains finite. In this limit, collective observables can approach classical behavior even though finite local quantum corrections persist at every site. The coexistence of these two limits demonstrates that classical behavior in an extended magnetic system can emerge either from large local spin or from large system size, and these mechanisms need not produce identical microscopic dynamics.

This distinction is particularly important for atomistic simulations. Classicality of experimentally relevant observables does not necessarily imply that every local quantum degree of freedom has become classical. A theory that retains the latter while remaining computationally scalable can therefore capture effects that are invisible to conventional LL/LLG simulations without incurring the exponential cost of a fully entangled quantum treatment. Because the equations of motion involve only the local density operators and self-consistent mean fields, their computational cost scales as $\mathcal{O}(N)$ for fixed spin $s$. The resulting framework consequently occupies an intermediate regime between classical atomistic spin dynamics and exponentially hard fully quantum many-body dynamics.

The paper is organized as follows. In Sec.~\ref{sec:cd}, we briefly review the classical LL and LLG equations and identify the structural ingredients that motivate their quantum counterparts. Section \ref{sec:qd} introduces the product-state \textit{ansatz} and derive the local quantum Landau-Lifshitz-Gilbert ($q$-LLG) equation using the Dirac-Frenkel variational principle supplemented by a Rayleigh-Lagrange-d'Alembert construction, and derives the corresponding local quantum Landau-Lifshitz ($q$-LL) equation using an Onsager gradient-flow approach; we also formulate both equations in terms of generalized coherence vectors and clarify their inequivalence for mixed states. In Sec.~\ref{sec:numerical-simulations}, we present three benchmarks addressing first-order anti-coherence generation and the sources of nonlinearity, the relation between quantum and classical dynamics and the inequivalence of $q$-LL and $q$-LLG in the large-spin limit, and the emergence of macroscopic quantum-classical agreement with increasing system size despite finite local discrepancies. Finally, Sec.~\ref{sec:conclusions} summarizes our results and discusses the implications of the product-state framework.  

\section{Classical spin dynamics}
\label{sec:cd}

The classical LL and LLG equations provide the reference point for the quantum construction developed in this work. Here, we identify the essential structural ingredients of these equations that will be retained when the classical spin magnetizations are replaced by quantum density operators. Two features will be important. First, the local field acting on a spin is determined by the state of the other spins through the energy functional, which leads naturally to a local mean-field description and an effective spin-spin interaction. Second, the cross-product structure of the classical dynamics constrains the motion to surfaces of constant spin magnitude. In the quantum setting, these two properties will be replaced, respectively, by a local effective Hamiltonian obtained from the product-state energy and by motion on an isospectral orbit of the local density operator. This provides the geometric and dynamical connection between the classical LL/LLG framework and the local quantum equations derived below.

To see this in more detail, we briefly outline the classical LL and LLG dynamics for $N$ spin magnetizations ${\bf m}_k$ with energy function
$H=H({\bf m}_1,\ldots,{\bf m}_N)$. These two forms of dynamics are described by
\begin{eqnarray}
\dot{\bf m}_j = \tilde{\gamma}_g {\bf m}_j \times {\bf B}_j
-\frac{\lambda}{|{\bf m}_j|}
{\bf m}_j \times ({\bf m}_j \times {\bf B}_j),
\label{eq:LL}
\end{eqnarray}
and
\begin{eqnarray}
\dot{\bf m}_k = \gamma_g {\bf m}_k \times {\bf B}_k
-\frac{\alpha}{|{\bf m}_k|}
{\bf m}_k \times \dot{{\bf m}}_k ,
\label{eq:llg}
\end{eqnarray}
respectively. Here, $\gamma_g$ is the gyromagnetic ratio, $\alpha$ is the dimensionless Gilbert damping parameter, the rescaled gyromagnetic ratio is
$\tilde{\gamma}_g=\gamma_g/(1+\alpha^2)$, the LL damping rate is
$\lambda=\alpha\gamma_g/(1+\alpha^2)$, and
\begin{equation}
{\bf B}_j=-\frac{\partial H}{\partial {\bf m}_j}
\end{equation}
is the local magnetic field seen by ${\bf m}_j$. The cross-product structure of these equations implies
${\bf m}_j\cdot\dot{\bf m}_j=0$, and hence $|{\bf m}_j|$ is conserved. 
These observations motivate the two corresponding ingredients of the quantum construction: the local effective field will be replaced by a local effective Hamiltonian operator, while conservation of the classical spin magnitude will be replaced by conservation of the spectrum of the local density operator.

\section{Quantum spin dynamics}
\label{sec:qd}

\subsection{The product state \textit{ansatz} and derivation of the local $q$-LLG equation of motion}

We develop a quantum analog of the just described classical dynamics by using the \textit{ansatz}
\begin{equation}
\label{eq:product-state-ansatz}
\varrho(t)=\bigotimes_{j=1}^{N}\varrho_j(t). 
\end{equation}
Motivated by the fact that in Ref.~\cite{liu24} it was shown that the quantities $C_n={\rm Tr}\varrho^n$, for all $n\in\mathbb{N}$, are invariants of the $q$-LLG equation of motion and that, as a consequence, the spectrum is preserved in time, we restrict each local density operator to an isospectral orbit through $\varrho_j(0)$, namely
\begin{equation}
\label{eq:isospectral-orbit}
\mathscr{O}_j
=\left\{\mathcal{U}_j\varrho_j(0)\mathcal{U}_j^{\dagger}:\mathcal{U}_j\in{\rm SU}(d)\right\},
\end{equation}
where $\mathcal{U}_j = \mathcal{U}_j (t,\varrho(0))$ satisfies    $\mathcal{U}_j(0,\varrho(0)) = \mathbbm{1}$, and $d=2s+1$. In other words, the orbit is determined by the initial local spectrum, whereas the trajectory on that orbit depends on the complete initial product state $\varrho(0)$ through the coupled equations of motion. Accordingly, the propagator $\mathcal{U}_j$ is state-dependent and does not define a single unitary channel acting identically on all input states $\varrho_j(0)$. Keeping this in mind, we suppress the initial state argument of $\mathcal{U}_j$ for brevity \footnote{A natural extension of this framework is to include the current state $\varrho(t)$ as an argument of the propagator itself, setting $\mathcal{U}_j \equiv \mathcal{U}_j(t,  \varrho(t))$. In this scenario, the dynamics are governed by the following \textit{implicit} equation of motion:
\begin{equation}
\label{eq:new-implicit-dynamics}
(\mathbbm{1} - \mathscr{M})\dot{\varrho} = \frac{i}{\hbar} [\varrho, \mathscr{A}],
\end{equation}
where we define $\mathcal{U}(t, \varrho) \equiv \bigotimes_{j=1}^N \mathcal{U}_j(t, \varrho)$, the shorthand $\partial_\circ \equiv \partial/\partial\circ$, and the differential $D_\varrho \mathcal{U}$ with respect to $\varrho$. This allows us to introduce the operators $\mathscr{A} \equiv i\hbar (\partial_t \mathcal{U}) \mathcal{U}^\dagger$ and $\mathscr{M} \equiv i\hbar^{-1} [\varrho, i\hbar (D_\varrho \mathcal{U}) \mathcal{U}^\dagger]$. This formulation does not intrinsically introduce a memory kernel. Although implicit, the dynamics described by Eq.~(\ref{eq:new-implicit-dynamics}) remain Markovian since the evolution depends only on the instantaneous global state, under the hypotheses of regularity and uniqueness. However, provided that $(\mathbbm{1} - \mathscr{M})$ is invertible, it can compel $\varrho$ to explore distinct paths in state space, thus inducing a qualitatively different relaxation process. More generally, this class of equations can be uniquely expressed in the form $\dot{\varrho} = \mathcal{F}(t, \varrho)$, where $\mathcal{F}$ is a vector field on the manifold $\mathcal{M}$. The specific choice of $\mathcal{U}$ alters $\mathcal{F}$, modifying its solutions and stability properties, but it does not introduce new degrees of freedom.}. Such unitary transformation preserves ${\rm Tr}\varrho_j^m$ ($m=1,\ldots,d$), and therefore the  spectrum, Hermiticity, and positivity of the local state $\varrho_j$ during dynamics; this is the quantum analog of preserving the magnitude of the classical spin vector, which reduces to fixing the Bloch-vector magnitude in the case of $s=\frac{1}{2}$. The dynamics is thus restricted to the manifold
\begin{equation}
\label{eq:manifold-definition}
\mathcal{M}=\prod_{j=1}^N\mathscr{O}_j.
\end{equation}

Next, we show that the conservative part of the dynamics follows from the Dirac-Frenkel time-dependent variational principle (DF-TDVP) \cite{kramer81}. The DF-TDVP states that, for a normalized trial wave function $\Psi$ depending on variational parameters $\tilde{\lambda}_j(t)$,
\begin{equation}
\label{eq:dirac-frenkel}
\langle\delta\Psi|\left(i\hbar\frac{d}{d t}-\mathscr{H}\right)|\Psi\rangle=0,
\end{equation}
for a given Hamiltonian $\mathscr{H}$ and for all variations $|\delta\Psi\rangle$ in the tangent space spanned by the derivatives $\partial|\Psi\rangle/\partial\tilde{\lambda}_j$. Eq. (\ref{eq:dirac-frenkel}) corresponds to the principle of least action $\delta\mathcal{S}=0$, where the action $\mathcal{S}$ is
\begin{equation}
\label{eq:action}
\mathcal{S}=\int dt\left[i\hbar\langle\Psi|\dot{\Psi}\rangle-\underbrace{\langle\Psi|\mathscr{H}|\Psi\rangle}_{E}\right],
\end{equation}
in which the first and second terms are, respectively, the kinematic term and the variational energy $E$. Although Eqs.~(\ref{eq:dirac-frenkel}) and (\ref{eq:action}) express the Dirac-Frenkel variational principle in its  familiar pure state form, the dynamics described by Eq. (\ref{eq:isospectral-orbit}) does not require the local density operators to be pure. This apparent discrepancy can be resolved by means of a purification scheme, as detailed in Appendix \ref{sec:purification-scheme}. For both pure and mixed local states, we can thus express the action $\mathcal{S}$ as
\begin{equation}
\label{eq:orbit-action-general}
\mathcal{S}[\{\mathcal{U}_j\}]=\int dt\left\{\sum_j i\hbar{\rm Tr}_{j}\left[\varrho_j(0)\mathcal{U}_j^{\dagger}\dot{\mathcal{U}}_j\right]-E[\{\varrho_j\}]\right\}.
\end{equation}

Given Eq. (\ref{eq:orbit-action-general}), we next evaluate the least-action condition $\delta\mathcal{S}=0$. Since the variation $\delta E$ is the sum of all independent first-order variations in each factor $j$, i.e., $\delta E = \sum_j \delta_j E$, with $\delta_jE={\rm Tr}\left[\left(\varrho_1\otimes\cdots\otimes\delta\varrho_j\otimes\cdots\otimes\varrho_N\right)\mathscr{H}\right]$ \footnote{Let $\varrho_j(\epsilon)=\varrho_j+\epsilon\delta\varrho_j$. By multilinearity, $E(\epsilon)={\rm Tr}\{[\bigotimes_{j=1}^N\varrho_j(\epsilon)]\mathscr{H}\}=E^{(0)}+\epsilon\sum_{j=1}^N{\rm Tr}\{(\varrho_1\otimes\cdots\otimes\delta\varrho_j\otimes\cdots\otimes\varrho_N)\mathscr{H}\}+\mathcal{O}(\epsilon^2)$. Hence $\delta E=\left.dE/d\epsilon\right|_{\epsilon=0}=\sum_j\delta_jE$.}, by defining $\mathcal{H}_{\bar{j}}\equiv\bigotimes_{k\neq j}\mathcal{H}_k$ as the Hilbert space of all sites except $j$, and $\varrho_{\bar{j}}\equiv\bigotimes_{k\neq j}\varrho_k$, one can write
\begin{equation}
\label{eq:delta-j-E}
\delta_j E={\rm Tr}\left[\left(\delta\varrho_j\otimes\varrho_{\bar{j}}\right)\mathscr{H}\right].
\end{equation}
Equation (\ref{eq:delta-j-E}) can be rearranged using the defining property of the partial trace, ${\rm Tr}_{j\bar{j}}\left[\left(\mathcal{A}_j\otimes\mathbbm{1}_{\bar{j}}\right)X\right]={\rm Tr}_{j}\left[\mathcal{A}_j{\rm Tr}_{\bar{j}}X\right]$, to obtain
\begin{equation}
\label{eq:delta-j-E-rearranged}
\delta_jE={\rm Tr}_j\left\{\delta\varrho_j\underbrace{{\rm Tr}_{\bar{j}}\left[\left(\mathbbm{1}_{j}\otimes\varrho_{\bar{j}}\right)\mathscr{H}\right]}_{h_j}\right\}={\rm Tr}_j\left(\delta\varrho_jh_j\right),
\end{equation}
from which we define $h_j$ as the \textit{local effective mean-field Hamiltonian}, and thus $\delta E =\sum_j{\rm Tr}_j\left(\delta\varrho_jh_j\right)$. Terms proportional to identity may be omitted from $h_j$, since they do not contribute to either equation of motion.

To completely determine the conservative part of equation of motion from the principle of least action, we need to show that, since $\varrho_j(t)$ is described as an isospectral orbit around $\varrho_j(0)$, both the local tangent variation $\delta\varrho_j$ and the tangent velocity $\dot{\varrho}_j$ in state space can be expressed as commutators with Hermitian operators $G_j$ and $A_j$. 
For the tangent velocity $\dot{\varrho}_j$, given that $\dot{\varrho}_j=\dot{\mathcal{U}}_j\varrho_j(0)\mathcal{U}_j^{\dagger}+\mathcal{U}_j\varrho_j(0)\dot{\mathcal{U}}_j^{\dagger}$, we define
\begin{equation}
\label{eq:definition-aj}
A_j=i\hbar\dot{\mathcal{U}}_j\mathcal{U}_j^{\dagger} , 
\end{equation}
which is Hermitian by unitarity of $\mathcal{U}_j$, and obtain $\dot{\varrho}_j=(i/\hbar)[\varrho_j,A_j]$. In turn, for the local variation $\delta\varrho_j = \delta \mathcal{U}_j \varrho_j (0) \mathcal{U}_j^{\dagger} + \mathcal{U}_j \varrho_j (0) \delta \mathcal{U}_j^{\dagger}$, let a unitary infinitesimal variation of $\mathcal{U}_j$ be $\mathcal{U}_j\rightarrow e^{-iG_j/\hbar}\mathcal{U}_j$ with $G_j^{\dagger}=G_j$. Then,  
\begin{equation}
\label{eq:definition-gj}
\delta\mathcal{U}_j=-\frac{i}{\hbar}G_j\mathcal{U}_j , 
\end{equation}
which gives $\delta\varrho_j=(i/\hbar)[\varrho_j,G_j]$. From Eqs. (\ref{eq:definition-aj}) and (\ref{eq:definition-gj}) we can write a useful relation between $A_j$ and $G_j$:
\begin{equation}
\label{eq:relation-aj-gj}
\delta A_j=\dot{G}_j+\frac{i}{\hbar}\left[A_j,G_j\right],
\end{equation}
which is precisely the quantum analog of the variation of the generalized velocity in classical mechanics. Indeed, for a classical generalized coordinate $q_j$, one has $\delta \dot q_j=(d/dt)\delta q_j$, so that the correspondence $\dot q_j\leftrightarrow A_j$ and $\delta q_j\leftrightarrow G_j$ can be made. In Eq.~(\ref{eq:relation-aj-gj}), however, the extra commutator term $[A_j,G_j]$ appears because $\mathcal U_j(t)$ evolves in the non-Abelian group ${\rm SU}(d)$.

From Eq.~\eqref{eq:orbit-action-general}, after integrating the resulting term $\sum_j\int dt\operatorname{Tr}_j(\varrho_j\dot G_j)$ by parts assuming that $G_j$ vanishes at the temporal endpoints, and using the relationships given in Eqs. (\ref{eq:delta-j-E-rearranged})-(\ref{eq:relation-aj-gj}), the stationarity condition becomes
\begin{equation}
\label{eq:variation-action}
\delta\mathcal{S}=-\sum_j\int dt{\rm Tr}_j\left\{G_j\left(\dot{\varrho}_j+\frac{i}{\hbar}[h_j,\varrho_j]\right)\right\}=0,
\end{equation}
which gives
\begin{equation}
\label{eq:conservative-part}
\dot{\varrho}_j=\frac{i}{\hbar}[\varrho_j,h_j].
\end{equation}
Here, a nonlinearity of the dynamics naturally emerges from the local mean-field Hamiltonian, since $h_j$ depends on all other density operators $\varrho_k$ ($k\neq j$) that are coupled to site $j$.

The DF-TDVP itself cannot fix the non-conservative part of the equation of motion. To this aim, dissipation can be incorporated through a Lagrange-d'Alembert (LA) extension of the variational principle, in which the conservative TDVP condition  $\delta\mathcal{S}=0$ is replaced by
\begin{equation}
\label{eq:dalembert-extension}
\delta\mathcal{S}+\int dt\delta\mathcal{W}_{\rm diss}=0,
\end{equation}
where $\delta\mathcal{W}_{\rm diss}$ denotes the work of the dissipative generalized force under an infinitesimal virtual displacement, whose time integral balances the variation of the action. Thus, a Gilbert-like dissipative force can be generated by the following Rayleigh function $\mathcal{R}$, which, in analogy to classical mechanics, where it is proportional to the square of the velocity of each particle in a given system, can be written in the form
\begin{equation}
\label{eq:rayleigh-dissipation-function}
\mathcal{R}=\frac{1}{2}\sum_j\eta_j{\rm Tr}_j\dot{\varrho}_j^2,
\end{equation}
since $\dot{\varrho}_j$ is the local tangent velocity in state space and where  $\eta_j>0$ are time-independent parameters \footnote{Here, constant coefficients $\eta_j$ are assumed for simplicity, but this is not \textit{a priori} required by the variational construction. More generally, let $\eta_j\equiv \eta_j(\{\varrho_k\},t)>0$. Since the Rayleigh dissipative force terms are derived via $\partial\mathcal{R}/\partial\dot{\varrho}_j$, $\delta\mathcal{W}_{\rm diss}$ will retain the exact form of Eq.~(\ref{eq:virtual-work1}) in the main text even in this broader scenario. As a consequence, the same \textit{q}-LLG equation [Eq. (\ref{eq:local-llg})] is preserved by setting $\kappa_j\equiv\eta_j/\hbar$, meaning that a complex dependence of $\eta_j$ may be simply absorbed into the damping parameter $\kappa_j$. A further generalization includes couplings between velocities at different sites in the Rayleigh functional, as in classical LLG models with spatially nonlocal Gilbert damping~\cite{Lu2023}.}\cite{Lu2023}. The dissipative generalized force is defined by the derivative of $\mathcal{R}$ with respect to the generalized velocity. In the present density-operator manifold, we thus have:
\begin{equation}
\label{eq:virtual-work1}
\delta\mathcal{W}_{\rm diss}=-\sum_j{\rm Tr}_j\left(\frac{\partial\mathcal{R}}{\partial\dot{\varrho}_j}\delta\varrho_j\right)=-\sum_j\eta_j{\rm Tr}_j\left(\dot{\varrho}_j\delta\varrho_j\right).
\end{equation}
Replacing $\delta\varrho_j = (i/\hbar) [\varrho_j,G_j]$ into Eq. (\ref{eq:virtual-work1}), and using Eqs. (\ref{eq:dalembert-extension}) and (\ref{eq:variation-action}), we arrive at:
\begin{equation}
\label{eq:variation-action2}
\begin{gathered}
\delta\mathcal{S}+\int dt\delta\mathcal{W}_{\rm diss}=\\
-\sum_j\int dt{\rm Tr}_j\left\{G_j\left(\dot{\varrho}_j+\frac{i}{\hbar}[h_j,\varrho_j]+\frac{i\eta_j}{\hbar}[\dot{\varrho}_j,\varrho_j]\right)\right\}=0,
\end{gathered}
\end{equation}
where, upon defining $\kappa_j\equiv\eta_j/\hbar>0$ and using the arbitrariness of $G_j$, one finally obtains:
\begin{equation}
\label{eq:local-llg}
\dot{\varrho}_j=\frac{i}{\hbar}[\varrho_j,h_j]+i\kappa_j[\varrho_j,\dot{\varrho}_j].
\end{equation}

Equation~(\ref{eq:local-llg}) is the local $q$-LLG equation on $\mathcal{M}$. It is \textit{local} in the sense that the dynamical variable is the single-site density operator $\varrho_j$, but \textit{nonlinear} due to two distinct origins. First, for interacting sites, imposing the product-state \textit{ansatz} [Eq.~(\ref{eq:product-state-ansatz})] throughout the evolution gives effective mean-field Hamiltonians $h_j$ that depend on the instantaneous states of other sites; the coupled conservative equations [Eq.~(\ref{eq:conservative-part})] are thus nonlinear even in the absence of damping, with the nonlinearity resulting from the restriction of the interacting dynamics to the manifold $\mathcal{M}$ itself. Second, the dissipative term, $i\kappa_j[\varrho_j,\dot{\varrho}_j]$,  introduces a nonlinear dependence on the local state $\varrho_j$, even in the case of a single site with a non-interacting 
Hamiltonian. 
For a time-independent Hamiltonian $\mathscr{H}$, the dissipative character of Eq.~(\ref{eq:local-llg}) follows directly from the Rayleigh construction, since $dE/dt=\sum_j{\rm Tr}_j(h_j\dot{\varrho}_j)=-\sum_j\hbar\kappa_j{\rm Tr}_j(\dot{\varrho}_j^2)\leq0$ \footnote{Equation (\ref{eq:local-llg}) of the main text can be written as $ \dot{\varrho}_j=(i/\hbar)[\varrho_j,h_j+\hbar\kappa_j\dot{\varrho}_j]$. Multiplying both sides by $h_j+\hbar\kappa_j\dot{\varrho}_j$ and taking the trace yields ${\rm Tr}_j(h_j\dot{\varrho}_j)
+\hbar\kappa_j{\rm Tr}_j(\dot{\varrho}_j^{,2})=0$ by cyclicity of the trace. Summing over sites gives the $dE/dt$ expression. The inequality follows from Hermiticity of $\dot{\varrho}_j$ and $\kappa_j\geq0$.}. In this formulation, the environmental degrees of freedom responsible for damping are not treated explicitly; instead, their effect is represented by a Rayleigh-type virtual work term, in direct analogy with the phenomenological Gilbert damping used in classical spin dynamics \cite{gilbert04}.

\subsection{Derivation of the local $q$-LL equation of motion}

The description of quantum spin dynamics has not only one form -- the $q$-LLG -- but can also be represented by its corresponding Landau-Lifshitz form -- the $q$-LL --, introduced in Ref.~\cite{wieser13}. Unlike the classical case, in which both LL and LLG forms are equivalent up to a time rescaling \cite{lakshmanan84}, they are known to be generally inequivalent for nonpure states \cite{liu24} -- a distinct nonclassical feature of quantum spin dynamics. The predictions of $q$-LL and $q$-LLG, however, appear to be qualitatively similar for some numerical applications \cite{liu26}. Since, as mentioned, the DF-TDVP does  not fix the non-conservative part of the equation of motion, we can investigate how to derive the local $q$-LL equation on the same manifold of product states. 

To this aim, instead of a LA extension of the variational principle [Eq. (\ref{eq:dalembert-extension})], the $q$-LL equation can be obtained from an Onsager gradient flow construction on the same isospectral orbit. In contrast to the Rayleigh construction used before, where dissipation enters as a generalized force linear in the tangent velocity $\dot{\varrho}_j$, the Onsager principle of least dissipation \cite{Onsager1931} determines the dissipative velocities themselves by minimizing the sum of the instantaneous energy rate and a quadratic dissipation cost, where the slow variable is chosen to be $\varrho_j$.

In this sense, we decompose the velocity $\dot{\varrho}_j$ in a Hamiltonian ($\dot{\varrho}_j^{\rm H}$) and a dissipative ($\dot{\varrho}_j^{\rm D}$) part as:
\begin{equation}
\label{eq:varrho-decomp}
\dot{\varrho}_j=\dot{\varrho}_j^{\rm H}+\dot{\varrho}_j^{\rm D},
\end{equation}
where $\dot{\varrho}_j^{\rm H}=(i/\hbar)[\varrho_j,h_j]$ was already fixed by the DF-TDVP. In turn, since the dynamics is restricted to the local isospectral orbit, the velocity $\dot{\varrho}_j$ -- and thus $\dot{\varrho}_j^{\rm D}$ -- must, by construction, be tangent to this orbit:
\begin{equation}
\label{eq:dissipative-velocity}
\dot{\varrho}_j^{\rm D}=i[\varrho_j,\mathcal{K}_j],
\end{equation}
for a given Hermitian generator $\mathcal{K}_j=\mathcal{K}_j^{\dagger}$ \footnote{Every tangent velocity on a unitary orbit admits a Hermitian generator $\mathcal{K}_j$; this choice ensures that $i[\varrho_j,\mathcal{K}_j]$ is Hermitian.}. Up to here, $\mathcal{K}_j$ is not assumed to have any particular form, and therefore parametrizes all admissible dissipative velocities $\dot{\varrho}_j^{\rm D}$ -- including the term that gives origin to the local $q$-LLG [Eq. (\ref{eq:local-llg})], by choosing $\mathcal{K}_j=\kappa_j\dot{\varrho}_j$. Here, however, the generator $\mathcal{K}_j$ will be fixed by minimization of the following Onsagerian functional:
\begin{equation}
\label{eq:onsagerian}
\mathcal{O}[\{\mathcal{K}_j\}]=\frac{dE}{dt}\Bigg|_{\rm D}+\Phi,
\end{equation}
where the first term is the instantaneous energy rate of change associated with the dissipative part, and the second term $\Phi$ is a quadratic dissipation cost functional \cite{Onsager1931}. A minimal choice for this function which is quadratic, positive, additive over sites, and isotropic in the local ${\rm SU}(d)$ generators is
\begin{equation}
\label{eq:definition-of-Phi}
\Phi=\frac{1}{2}\sum_j\frac{1}{\tilde{\lambda}_j}{\rm Tr}_j(\mathcal{K}_j^2),
\end{equation}
where $\tilde{\lambda}_j>0$ are time-independent parameters \footnote{The generator is not unique: any $\mathcal{K}_j+\mathcal{Z}_j$ produces the same $\dot{\varrho}_j^{\rm D}$ whenever $[\mathcal{Z}_j,\varrho_j]=0$, with $\mathcal{Z}_j^{\dagger}=\mathcal{Z}_j$. We therefore minimize the quadratic cost $\Phi$ over this freedom. Any $\mathcal{K}_j$ can be decomposed as $\mathcal{K}_j=\mathcal{K}_j^{\parallel}+\mathcal{K}_j^{\perp}$, where $\mathcal{K}_j^{\parallel}=\sum_\alpha\mathcal{P}_{j,\alpha}\mathcal{K}_j\mathcal{P}_{j,\alpha}$, $\mathcal{K}_j^{\perp}=\sum_{\alpha\neq \beta}\mathcal{P}_{j,\alpha}\mathcal{K}_j\mathcal{P}_{j,\beta}$, and $\varrho_j\equiv\sum_\alpha p_{j,\alpha}\mathcal{P}_{j,\alpha}$. Thus, ${\rm Tr}(\mathcal{K}_j^{\perp}\mathcal{Z}_j)={\rm Tr}(\mathcal{K}_j^{\perp}\mathcal{K}_j^{\parallel})=0$, and ${\rm Tr}[(\mathcal{K}_j+\mathcal{Z}_j)^2]={\rm Tr}[(\mathcal{K}_j^{\perp})^2]+{\rm Tr}[(\mathcal{K}_j^{\parallel}+\mathcal{Z}_j)^2]$. Since the dissipative velocity and the energy rate $dE/dt$ depend only on $\mathcal{K}_j^{\perp}$, minimizing the dissipative cost implies that the optimal operator is $\mathcal{Z}_j^{\star}=-\mathcal{K}_j^{\parallel}$. Thus, the minimum-cost representative is $\tilde{\mathcal{K}}_j=\mathcal{K}_j^{\perp}$. It can be expressed as a commutator with $\varrho_j$, $\tilde{\mathcal{K}}_j\equiv i[\varrho_j,\mathcal{B}_j]$, with off-diagonal blocks $(\mathcal{B}_j)_{\alpha\beta}=-i(\mathcal{K}_j^\perp)_{\alpha\beta}/(p_{j,\alpha}-p_{j,\beta})$ -- which satisfy $\mathcal{B}_j=\mathcal{B}_j^{\dagger}$.}. Therefore, Eq. (\ref{eq:definition-of-Phi}) naturally represents the linear response cost of generating a tangent motion $\dot{\varrho}_j^{\rm D}$ \footnote{More general choices, for instance involving anisotropic terms such as 
\begin{equation}
\Phi=\frac{1}{2}\sum_{jk,ab}\Gamma_{jk,ab}\mathcal{K}_{j,a}\mathcal{K}_{k,b},
\end{equation}
where $\Gamma_{jk,ab}=\Gamma_{kj,ba}$ and $\mathcal{K}_j=\mathcal{K}_{j,a}\Lambda_a$, are possible, but Eq. (\ref{eq:definition-of-Phi}) is the minimal choice that leads to the double-bracket $q$-LL damping and guarantees a monotonic decay of the variational energy.}. Since $dE/dt\big|_{\rm D}=\sum_j{\rm Tr}_j(h_j\dot{\varrho}_j^{\rm D})=\sum_ji{\rm Tr}_j\left([h_j,\varrho_j]\mathcal{K}_j\right)$, the minimization of the Onsagerian functional with respect to $\mathcal{K}_j$ leads to the variation
\begin{equation}
\label{eq:var-onsagerian}
\delta\mathcal{O}=\sum_j{\rm Tr}_j\left[\left(  i[h_j,\varrho_j]+\frac{1}{\tilde{\lambda}_j}\mathcal{K}_j\right)\delta\mathcal{K}_j\right]=0,  
\end{equation}
which, given that $\delta\mathcal{K}_j$ is arbitrary in the relevant tangent subspace, reduces to the minimum condition:
\begin{equation}
\label{eq:minimum-condition}
\mathcal{K}_j=i\tilde{\lambda}_j[\varrho_j,h_j].
\end{equation}
Thus, from Eq. (\ref{eq:dissipative-velocity}), $\dot{\varrho}_j^{\rm D}=-\tilde{\lambda}_j[\varrho_j,[\varrho_j,h_j]]$, and we finally obtain the local $q$-LL equation on the manifold of product states:
\begin{equation}
\label{eq:q-ll-local}
\dot{\varrho}_j=\frac{i}{\hbar}[\varrho_j,h_j]-\tilde{\lambda}_j[\varrho_j,[\varrho_j,h_j]] . 
\end{equation}

\subsection{Coherence vector representations}
Both Eqs.~\eqref{eq:local-llg} and \eqref{eq:q-ll-local} can be readily expressed in terms of generalized coherence vectors $\vec{x}_j=(x_1,\ldots,x_{d^2-1})$, $d=2s+1,$ associated with the local density matrices $\varrho_j$. Following the standard parametrization \cite{bertlmann08}, we write, in the Einstein summation convention,
\begin{equation}
\label{eq:form-of-rho-j}
\varrho_{j}(t)=\frac{1}{d}\mathbbm{1}+x_{j,a}(t)\Lambda_a ,
\end{equation}
where
\begin{equation}
x_{j,a}=\frac{1}{2}{\rm Tr}(\varrho_j\Lambda_a).
\end{equation}
Here, $\Lambda_a$ are the traceless generators of the Lie algebra $\mathfrak{su}(d)$, satisfying
\begin{equation}
\label{eq:algebra1}
[\Lambda_a,\Lambda_b]=2i f_{abc}\Lambda_c,
\end{equation}
where $f_{abc}$ are the totally antisymmetric structure constants. We adopt the normalization
\begin{equation}
\label{eq:normalization-trace-lambda}
{\rm Tr}(\Lambda_a\Lambda_b)=2\delta_{ab}.
\end{equation}

To recast the equations of motion in this representation, we expand the local effective Hamiltonian as
\begin{equation}
h_j=h_{j,0}\mathbbm{1}+h_{j,a}\Lambda_a
=h_{j,0}\mathbbm{1}+\vec{h}_j\cdot\vec{\Lambda},
\end{equation}
where $\vec{\Lambda}=(\Lambda_1,\ldots,\Lambda_{d^2-1})$. We further introduce the \emph{Lie-algebra cross product},
\begin{equation}
\label{eq:generalized-cross-product}
(\vec{x}_j\wedge\vec{h}_j)_a
=f_{abc}x_{j,b}h_{j,c}.
\end{equation}
With this definition, the commutator between the local density matrix and the effective Hamiltonian takes the form
\begin{equation}
[\varrho_j,h_j]
=2i f_{abc}x_{j,b}h_{j,c}\Lambda_a
=2i(\vec{x}_j \wedge \vec{h}_j)_a \Lambda_a.
\end{equation}
Thus, the commutator is represented by the Lie-algebra cross product in the coherence-vector space, in direct analogy with the usual vector cross product for $\mathfrak{su}(2)$.

Since the energy satisfies
\begin{equation}
\frac{\partial E}{\partial x_{j,a}}=2h_{j,a},
\end{equation}
it is natural to introduce an \emph{effective generalized field} $\vec{B}_j\in\mathbb{R}^{d^2-1}$, with components
\begin{equation}
\label{eq:generalized-field}
B_{j,a}=-\frac{\partial E}{\partial x_{j,a}}=-2h_{j,a}.
\end{equation}
This field plays the role of a local effective field acting on the generalized coherence vector at site $j$.

In terms of $\vec{x}_j$ and $\vec{B}_j$, the conservative part of the quantum dynamics takes the precessional form \footnote{Here, $\vec{B}_j$ is defined in energy units, as follows from Eq. (\ref{eq:generalized-field}), and the prefactor $\hbar^{-1}$ is therefore dimensioned accordingly. If instead $\vec{B}_j$ is expressed in terms of a physical magnetic field measured in tesla, the corresponding precessional coefficient has the dimensions of a gyromagnetic ratio}
\begin{equation}
\dot{\vec{x}}_j = \hbar^{-1} \vec{x}_j \wedge \vec{B}_j.
\end{equation}
Consequently, the local $q$-LL equation [Eq.\eqref{eq:q-ll-local}] becomes a Landau-Lifshitz-type equation for the generalized coherence vector,
\begin{equation}
\label{eq:q-ll-coherence-representation}
\dot{\vec{x}}_j
=\hbar^{-1}\vec{x}_j\wedge\vec{B}_j
-\tilde{\lambda}_j^{\prime}\vec{x}_j\wedge
(\vec{x}_j\wedge\vec{B}_j),
\end{equation}
where $\tilde{\lambda}_j^{\prime}=2\tilde{\lambda}_j$. Thus, the double-bracket structure is mapped onto a generalized double cross-product structure in the coherence vector space.

The same representation can be applied to the local $q$-LLG equation. In terms of the generalized coherence vector, it takes the form
\begin{equation}
\label{eq:q-llg-coherence-representation}
\dot{\vec{x}}_j
=\hbar^{-1}\vec{x}_j\wedge\vec{B}_j
-\kappa_j^{\prime}\vec{x}_j\wedge\dot{\vec{x}}_j,
\end{equation}
where $\kappa_j^{\prime}=2\kappa_j$. Hence, both the $q$-LL and $q$-LLG dynamics admit a unified Landau-Lifshitz-type representation in the generalized coherence-vector space, with the ordinary vector cross product replaced by the Lie-algebra cross product defined in Eq.~\eqref{eq:generalized-cross-product}.

For $d=2$, the structure constants reduce to those of $\mathfrak{su}(2)$, and Eqs.~\eqref{eq:q-ll-coherence-representation} and \eqref{eq:q-llg-coherence-representation} reduce exactly to the classical Landau-Lifshitz and Landau-Lifshitz-Gilbert equations, respectively, for a Bloch vector of fixed magnitude. More generally, the total antisymmetry of $f_{abc}$ implies
\begin{equation}
\vec{x}_j\cdot\dot{\vec{x}}_j=0,
\end{equation}
in both cases, and hence
\begin{equation}
\frac{d}{dt}|\vec{x}_j|^2=0.
\end{equation}
Therefore, the Euclidean norm $|\vec{x}_j|$ is conserved, in analogy with the conservation of $|\mathbf{m}_j|$ in the corresponding classical dynamics. With the normalization in Eq. (\ref{eq:normalization-trace-lambda}), ${\rm Tr}(\varrho_j^2)=1/d+2|\vec{x}_j|^2$.

\subsection{On the inequivalence of the local $q$-LL and $q$-LLG: the variational and geometric origin}
\label{sec:inequivalence}

Although the general inequivalence between $q$-LL and $q$-LLG has been established \cite{liu24,liu26}, little has been discussed about its origin. The two local equations obtained above have a common source of the conservative part but they correspond to distinct variational dissipative mechanisms on the same local isospectral orbit. This distinction can be made clearer by introducing, at each site $j$, the linear superoperator $\mathcal{L}_j(X)=i[\varrho_j,X]$, which maps Hermitian operators $X$ to tangent vectors of the orbit -- in a construction similar to that of Eq. (\ref{eq:dissipative-velocity}). With this notation, the conservative TDVP term, for instance, reads $\dot{\varrho}_j^{\rm H}=\hbar^{-1}\mathcal{L}_j(h_j)$.

The Onsager construction gives a gradient flow correction on the orbit. In this case, the local $q$-LL equation can be written as ($\tilde{\lambda}_j=\kappa_j/\hbar$):
\begin{equation}
\label{eq:local-q-ll-again}
\dot{\varrho}_j=\frac{1}{\hbar}\left[\mathcal{L}_j+{\kappa_j}\mathcal{L}_j^2\right](h_j).
\end{equation}
The dissipative operator $\mathcal{L}_j^2(\cdot)$ is self-adjoint and negative semidefinite: the damping term, proportional to $\mathcal{L}_j^2(h_j)$, is the projected energy gradient flow associated with the Onsager cost function chosen for the dissipative generator. 

In turn, the local $q$-LLG has a different structure. Expressed in terms of $\mathcal{L}_j(\cdot)$, it reads 
\footnote{To see this, take the commutator with $\varrho_j$ of both sides of Eq.~\eqref{eq:local-llg}, yielding 
\begin{equation}
[\varrho_j,\dot{\varrho}_j] = \frac{i}{\hbar} [\varrho_j,[\varrho_j,h_j]] + 
i\kappa_j [\varrho_j,[\varrho_j,\dot{\varrho}_j]] . 
\end{equation}
Inserting back into the right-hand side of Eq.~\eqref{eq:local-llg} and rearranging terms, one obtains 
\begin{equation}
\dot{\rho}_j + \kappa_j^2 [\varrho_j,[\varrho_j,\dot{\varrho}_j]] = 
\frac{i}{\hbar} [\varrho_j,h_j] - \frac{\kappa_j}{\hbar} [\varrho_j,[\varrho_j,h_j]]  . 
\end{equation}
This can be rewritten in terms of $\mathcal{L}_j$ as 
\begin{equation}
(1-\kappa_j^2 \mathcal{L}_j^2) (\dot{\varrho}_j)  = 
\frac{1}{\hbar} \left( \mathcal{L}_j + \kappa_j \mathcal{L}_j^2 \right) (h_j)  
\end{equation}
since $\mathcal{L}_j^2 (X) = \mathcal{L}_j (i[\varrho_j,X]) = 
- [\varrho_j,[\varrho_j,X]]$. Note that $(1-\kappa_j^2 \mathcal{L}_j^2)^{-1}$ always exists since $\mathcal{L}_j^2$ is negative semidefinite. Thus, Eq.~\eqref{eq:local-q-llg-again} follows.}:
\begin{equation}
\label{eq:local-q-llg-again}
\dot{\varrho}_j=\frac{1}{\hbar} \left[(1-\kappa_j^2\mathcal{L}_j^2)^{-1}(\mathcal{L}_j+\kappa_j\mathcal{L}_j^2) \right] (h_j),
\end{equation}
and thus is not obtained by simply adding the same Onsager gradient flow term to the conservative, reversible part, but it actually involves a renormalization of both the reversible and dissipative parts by $(1-\kappa_j^2\mathcal{L}_j^2)^{-1}$. Whenever $\mathcal{L}_j^2(\cdot)$ acts as a scalar on the local tangent space, this renormalization reduces to a single factor relating the two local velocities $\dot{\varrho}_j^{q-{\rm LL}}$ and $\dot{\varrho}_j^{q-{\rm LLG}}$. This occurs for any non-maximally mixed spin-$1/2$ state and, more generally, for any local density operator with exactly two distinct eigenvalues. A common rescaling of time for the coupled system additionally requires the same factor at every evolving site. For a generic mixed state with $d=2s+1>2$, however, $\mathcal{L}_j^2(\cdot)$ is not proportional to the identity on the tangent space of the isospectral orbit: in the eigenbasis of $\varrho_j$, in which $\varrho_j={\rm diag}(p_{j,1},\ldots,p_{j,d})$, a tangent direction connecting eigenstates $a$ and $b$ with different populations has eigenvalue $-\Delta_{j,ab}^2$, where $\Delta_{j,ab}=p_{j,a}-p_{j,b}$; distinct gap magnitudes therefore produce different rescaling factors. 
The following theorem makes the criterion for the existence of a common time rescaling precise.

\textbf{Theorem} (\textit{criterion for a common time rescaling}): Let $\tilde{\lambda}_j=\kappa_j/\hbar$ and 
\begin{equation}
\varrho_j=\sum_a p_{j,a}\mathcal{P}_{j,a},
\end{equation}
where $a$ labels distinct eigenvalues and $\mathcal{P}_{j,a}$ projects onto the corresponding eigenspace. Evaluate both velocities at the same
product-state configuration, with the same effective Hamiltonians $h_j$. At a nonstationary configuration, a scalar $\zeta>0$ satisfies
\begin{equation}
\dot{\varrho}_j^{q-{\rm LL}}=\zeta\dot{\varrho}_j^{q-{\rm LLG}},\text{  }\forall j,
\end{equation}
if and only if
\begin{equation}
\label{eq:clock-def-theorem}
c_{j,ab}\equiv1+\kappa_j^2(p_{j,a}-p_{j,b})^2=\zeta 
\end{equation}
for every site $j$ and every active pair, defined by $a\neq b$ and $\mathcal{P}_{j,a}h_j\mathcal{P}_{j,b}\neq 0$. Here, $c_{j,ab}$ is the \textit{local clock} factor associated with the pair $(a,b)$, and $\zeta$ is the \textit{common clock factor} for the entire system. 

\textit{Proof.---} Let $\Delta_{j,ab}=p_{j,a}-p_{j,b}$. One can write:
\begin{equation}
\mathcal{P}_{j,a}\mathcal{L}_j(X)\mathcal{P}_{j,b}=i\Delta_{j,ab}\mathcal{P}_{j,a}X\mathcal{P}_{j,b},
\end{equation}
and, therefore, $\mathcal{P}_{j,a}\mathcal{L}_j^2(X)\mathcal{P}_{j,b}=-\Delta_{j,ab}^2\mathcal{P}_{j,a}X\mathcal{P}_{j,b}$. Substitution into the two equations of motion gives
\begin{equation}
\mathcal{P}_{j,a}\dot{\varrho}_j^{q-{\rm LL}}\mathcal{P}_{j,b}=\hbar^{-1}(i\Delta_{j,ab}-\kappa_j\Delta_{j,ab}^2)\mathcal{P}_{j,a}h_j\mathcal{P}_{j,b},
\end{equation}
and
\begin{equation}
\label{eq:imp-theorem}
\mathcal{P}_{j,a}\dot{\varrho}_j^{q-{\rm LLG}}\mathcal{P}_{j,b}=\frac{\mathcal{P}_{j,a}\dot{\varrho}_j^{q-{\rm LL}}\mathcal{P}_{j,b}}{1+\kappa_j^2\Delta_{j,ab}^2}.
\end{equation}
For every active pair, the numerator in Eq. (\ref{eq:imp-theorem}) is nonzero, so proportionality of the two velocities requires $c_{j,ab}=\zeta$. At the same time, if all active clock factors coincide with $\zeta$, the same proportionality holds in every block, since inactive blocks and blocks with $a=b$ have zero velocity. \QEDB

Some consequences of this theorem follow directly. For a local state with exactly two distinct eigenvalues, all tangent blocks have the same nonzero gap magnitude, so the local clock factor is independent of $h_j$. Pure local states have eigenvalues $1$ and $0$, so a common damping coefficient $\kappa_j\equiv\kappa$ gives $\zeta=1+\kappa^2$, independent of the spin magnitude or if the states are spin coherent. Mixedness alone does not directly imply inequivalence: a state proportional to a rank-$r_j$ projector, $\varrho_j=\mathcal P_j/r_j$ with $1<r_j<d$, also has a single nonzero gap magnitude and results in the local clock factor $\zeta_j=1+\kappa_j^2/r_j^2$. A common clock exists when these factors coincide and whenever $\kappa_j=\kappa$ and $r_j=r$ for all evolving sites $j$ in a given system.

The inequivalence of the local $q$-LL and $q$-LLG is therefore not a fundamental consequence of entanglement, but of the construction of two distinct dissipative laws on the same product-state manifold: the failure of $\mathcal{L}_j^2(\cdot)$ to act as a scalar on the local tangent directions driven by $h_j$ is precisely the variational and geometric origin of such inequivalence. In other words, while the Rayleigh route to $q$-LLG treats dissipation mechanically, by assigning a quadratic (Frobenius) cost to the actual tangent velocity $\dot{\varrho}_j$, the Onsager route to $q$-LL exposes a geometric structure: the variationally selected gradient flow is governed by a state-dependent metric on the isospectral orbit. In this sense, the spectral gaps $\Delta_{j,ab}$ that distinguish the allowed transition directions act as the effective metric data of this geometry. In such context, the $q$-LL construction is closer in spirit to information geometric metrics than to a ``flat" mechanical damping law.

\section{Numerical simulations}
\label{sec:numerical-simulations}

The numerical methods used in all three benchmarks are described in Appendix \ref{sec:numerical-methods}.

\subsection{Benchmark 1: first-order anti-coherence generation and sources of nonlinearity}
\label{sec:benchmark1}

\begin{figure*}[htb]
\centering
\makebox[\textwidth][c]{%
  \includegraphics[width=0.85\textwidth]{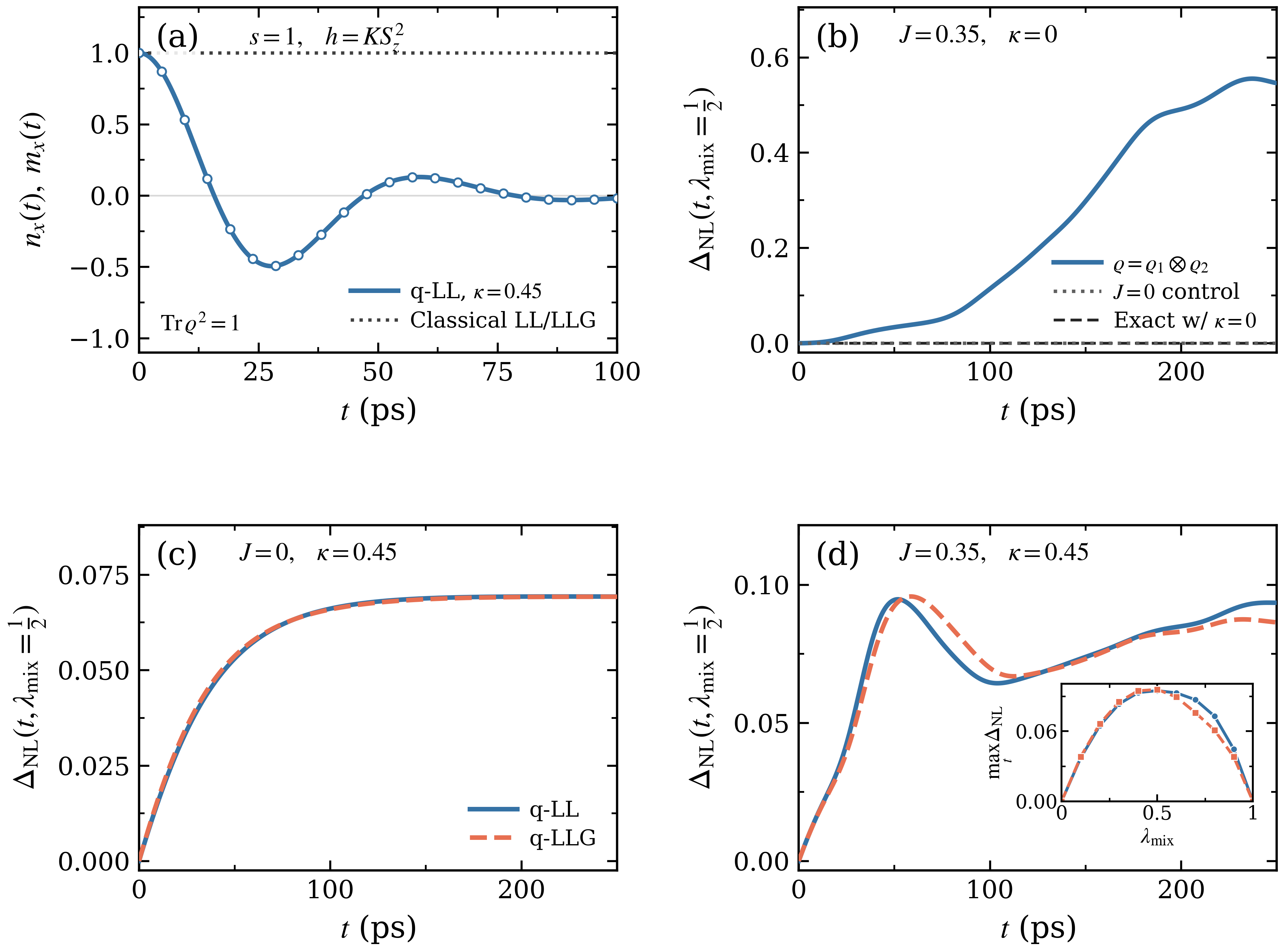}
}
\caption{\textbf{First-order anti-coherence generation and sources of nonlinearity.} (a) Dipole $x$ component, $n_x(t)$, of an initially coherent spin $s=1$ under $\mathscr{H}=E_0S_z^2$ for the $q$-LL evolution with $\kappa=0.45$. Lines denote analytical solutions (see Appendix \ref{sec:appendix-anticoherence}) and markers denote numerical results. The classical trajectory remains $\mathbf{m}(t)=\mathbf{e}_x$ since its initial torque vanishes. The first zero of the quantum dipole occurs at $t^{\star}=\pi\hbar/2E_0$, without loss of purity. (b)-(d) show the nonlinearity measure $\Delta_{\rm NL}$ for the two-spin preparation defined in the text, at $\lambda_{\rm mix}=\frac{1}{2}$: (b) $J\neq0$ and $\kappa=0$, (c) $J=0$ and $\kappa\neq 0$, and (d) both exchange and damping switched on. In panel (b), all curves have $\kappa=0$.  The dotted curve is the uncoupled limit ($J=0$), here used as a control, and the dashed curve denotes the exact unitary evolution of the full interacting two-spin density operator $\varrho(t)=e^{-i\mathscr{H}t/\hbar}\varrho(0)e^{i\mathscr{H}t/\hbar}$ -- where $\mathscr{H}$ is the complete Hamiltonian including the coupling between the spins. The inset in panel (d) shows the maximum $\Delta_{\rm NL}$ over the observation window as a function of $\lambda_{\rm mix}$. The local $q$-LL damping coefficient is set to $\tilde{\lambda}=\kappa/\hbar$ for comparison with the local $q$-LLG at damping $\kappa$. In panel (a), only the $x$ component is shown since $\mathbf{n}(t)\propto\mathbf{m}(t)=\mathbf{e}_x$ (see Appendix \ref{sec:appendix-anticoherence}).}
\label{fig:benchmark1}
\end{figure*}

We first evaluate two minimal models composed of $N=1$ and $N=2$ spins, respectively, designed to expose two features of the product-state dynamics: the possible selection of anti-coherent spin states, and sources of nonlinearity. Each spin is described by a local density matrix $\varrho_j(t) \in \mathbb{C}^{(2s+1)\times (2s+1)}$, with $j=1$ or $j=1,2$, evolving on its isospectral orbit [see Eq.~(\ref{eq:isospectral-orbit})]. Since for $s=\frac{1}{2}$ both the $q$-LL and the $q$-LLG dynamics become equivalent under a global time rescaling factor -- assuming that all sites have the same Bloch vector radius $|\vec{x}_j|$ and the same damping value $\kappa_j$ -- we here choose $s=1$ in both examples. In this case, the $j$-th spin operator $\mathbf{S}_j$ components are written in terms of the Gell-Mann matrices \cite{bertlmann08}:
\begin{equation}
\label{eq:gellmann} 
S_{x,j}=\frac{\lambda_1+\lambda_6}{\sqrt{2}},\quad
S_{y,j}=\frac{\lambda_2+\lambda_7}{\sqrt{2}},\quad
S_{z,j}=\frac{\lambda_3+\sqrt{3}\lambda_8}{2}.
\end{equation}

First, we consider a single spin with uniaxial anisotropy, described by the Hamiltonian
\begin{equation}
\label{eq:spin-1-ham-bench1}
\mathscr{H}=E_0K_zS_z^2,
\end{equation}
where $E_0=(\hbar/10\,{\mathrm{ps}})\approx 0.06582$~meV is a constant, and $K_z=1$. The system is initialized in the coherent state $|\psi(0)\rangle=(|1\rangle+\sqrt{2}|0\rangle+|-1\rangle)/2$ corresponding to the initial quantum dipole $\mathbf{n} (0) = \bra{\psi (0)} \mathbf{S} \ket{\psi (0)} = \mathbf{e}_x$. As derived in Appendix~\ref{sec:appendix-anticoherence}, for the $q$-LL equation of motion with $\tilde{\lambda}=\kappa/\hbar$, the quantum dipole evolves under $\mathscr{H}$ in Eq.~\eqref{eq:spin-1-ham-bench1} as
\begin{equation}
\langle\mathbf{S}\rangle=\mathbf{n}(\tau)=\frac{\cos\tau}{\cosh(\kappa\tau)}\mathbf{e}_x,
\end{equation}
where $\tau=E_0t/\hbar$ is the dimensionless time parameter, and the state remains pure throughout the evolution. As $\tau\to\infty$, $\mathbf{n} (\tau) \to 0$, asymptotically producing first-order anti-coherence \cite{zimba06,Baguette2017}.

Next, to separate the sources of nonlinear evolution, we consider a two-spin system with $s=1$. The system state is given by $\varrho(t)=\varrho_1(t)\otimes\varrho_2(t)$, and the spin Hamiltonian is
\begin{equation}
\label{eq:spin-hamiltonian-benchmark1}
\mathscr{H}=E_0\left\{\sum_{j=1}^2\left[-\mathbf{b}\cdot\mathbf{S}_j+K_xS_{x,j}^2+K_zS_{z,j}^2\right]-J\mathbf{S}_1\cdot\mathbf{S}_2\right\},
\end{equation}
which, under the product-state \textit{ansatz}, gives the following local mean-field Hamiltonian $h_j$, as defined in Eq.~(\ref{eq:delta-j-E-rearranged}):
\begin{equation}
\label{eq:local-hamiltonian-test-1}
h_j=E_0\left(-\mathbf{b}\cdot\mathbf{S}_j+K_xS_{x,j}^2+K_zS_{z,j}^2-J\mathbf{n}_k\cdot\mathbf{S}_j\right),\quad (k\neq j).
\end{equation}
In 
Eq.~(\ref{eq:local-hamiltonian-test-1}),
\begin{equation}
\label{eq:normalized-moment-quantum}
\mathbf{n}_j\equiv\frac{1}{s}{\mathrm{Tr}}(\varrho_j\mathbf{S}_j)
\end{equation}
is the normalized quantum dipole, $\mathbf{b}\equiv\mathbf{B}/|\mathbf{B}|$ with $\mathbf{B}=(1,3/20,1)$ is the normalized applied magnetic field vector specifying the dimensionless Zeeman coupling, $J$ is the magnitude of the exchange interaction, and $K_{x}=0.3$ and $K_z=-0.12$ are magnetic anisotropy parameters. The exchange and damping are switched between $J=0, 0.35$ and $\kappa=0, 0.45$ to isolate their contributions. All interaction parameters, including $K_x$, $K_z$, $J$, and the components of $\mathbf{b}$, are dimensionless; the prefactor $E_0$ sets the energy scale of the Hamiltonian. The anisotropy terms proportional to $S_x^2$ and $S_z^2$ are essential to this benchmark: they provide a nontrivial test of quantum-classical correspondence, since their contribution to the dipole precession depends on second-order spin moments, which are generally not determined by the dipole alone. They can therefore drive initially coherent local states outside the spin-coherent-state manifold.

In this two-spin part, the local states are initially prepared as
\begin{equation}
\label{eq:initial-local-states-bench1}
\varrho_j(0)=\mathfrak{R}(\theta_j,\varphi_j){\mathrm{diag}}(\mathbf{p}_j)\mathfrak{R}^{\dagger}(\theta_j,\varphi_j),
\end{equation}
where $\mathfrak{R}(\theta,\varphi)\equiv e^{-i\varphi S_z}e^{-i\theta S_y}$, $(\theta_j,\varphi_j)$ are sets of randomly assigned spherical angles, and $\mathbf{p}_j\equiv(p_{j,1},p_{j,2},p_{j,3})$ 
is a vector containing the three randomly assigned eigenvalues of $\varrho_j(0)$ such that $p_{j,a}\geq0$ and $\sum_ap_{j,a}={\mathrm{Tr}}[\varrho_j(0)]=1$.

To construct a measure of nonlinearity, we consider the following. Without loss of generality, we assign to site 1 two distinct preparations -- here denoted as $A$ and $B$ -- following the definition of Eq.~(\ref{eq:initial-local-states-bench1}); specifically, we define two distinct sets of random spherical angles and eigenvalues. In turn, site 2 is assigned a unique set of preparation parameters. We introduce a mixing parameter $0\leq\lambda_{\rm mix}\leq 1$ such that 
\begin{equation}
\label{eq:initial-rho1-bench1}
\varrho_1(0)=\lambda_{\rm mix}\varrho_{1,A}(0)+(1-\lambda_{\rm mix})\varrho_{1,B}(0),
\end{equation}
and $\varrho_2(0)$ is independent of $\lambda_{\rm mix}$. Thus, the initial state takes the form 
\begin{equation}
\label{eq:affine-identity-bench1}
\varrho(0)=\lambda_{\rm mix}[\varrho_{1,A}(0)\otimes\varrho_2(0)]+(1-\lambda_{\rm mix})[\varrho_{1,B}(0)\otimes\varrho_2(0)].
\end{equation}
For each quantum equation of motion, we evolve $A$, $B$, and their initial mixture independently, and evaluate
\begin{widetext}
\begin{equation}
\label{eq:nonlinearity-measure-bench1}
\Delta_{\rm NL}(t,\lambda_{\rm mix})=\left[\frac{1}{2}\sum_{j=1}^2{\mathrm{Tr}}\left\{[\varrho_j(t)-\lambda_{\rm mix}\varrho_{j,A}(t)-(1-\lambda_{\rm mix})\varrho_{j,B}(t)]^2\right\}\right]^{1/2},
\end{equation}
\end{widetext}
where, as defined previously, $\varrho_{2,A}(0)=\varrho_{2,B}(0)=\varrho_2(0)$. Equation~(\ref{eq:nonlinearity-measure-bench1}) thus combines the local Hilbert-Schmidt distances to quantify the difference between evolving the initial mixture and mixing the separately evolved local states.

Figure~\ref{fig:benchmark1} presents the results for both the single-spin and two-spin tests. As shown in Fig.~\ref{fig:benchmark1}(a) for a single $s=1$ spin, the nonzero damping asymptotically selects the first-order anti-coherent state with $\langle\mathbf{S}\rangle = n_x (t) \mathbf{e}_x = 0$. Figure~\ref{fig:benchmark1}(b) illustrates that when $J=\kappa=0$, the dynamics remains linear despite the quadratic anisotropy (i.e., $\Delta_{\rm NL}=0$); similarly, the exact conservative evolution ($\kappa = 0$)
of the interacting ($J\neq 0$) two-spin system
is also linear. In contrast, the undamped (conservative) product-state dynamics with $J\neq 0$ develops a nonzero $\Delta_{\rm NL}$, tracing the origin of such nonlinearity to the restriction of the dynamics to the manifold $\mathcal{M}$. In panel (c), where $J=0$ and $\kappa\neq 0$, the local Hamiltonians remain prescribed, yet both quantum dissipative equations produce a nonzero $\Delta_{\rm NL}$. Thus, even in a non-interacting scenario, the nonlinearity can also arise from the structure of the dissipative term itself. Panel (d) combines both contributions. By construction, $\Delta_{\rm NL}$ vanishes in all cases when $\lambda_{\rm mix}=0$ or $\lambda_{\rm mix}=1$; its behavior at intermediate values, shown in the inset of panel (d), tests the dependence on the initial preparation. Ultimately, Fig.~\ref{fig:benchmark1} demonstrates that the absence of intersite entanglement does not make the dynamics classical: a nonzero damping can, both asymptotically and for finite $t$, select a pure anti-coherent state that a single classical spin of fixed magnitude cannot represent. At the same time, the product-state equations do not necessarily preserve the linearity of the exact unitary evolution. 

 
\subsection{Benchmark 2: the large spin limit and $q$-LL/$q$-LLG inequivalence}
\label{sec:benchmark2}

\begin{figure*}[htb]
\centering
\makebox[\textwidth][c]{%
  \includegraphics[width=0.9\textwidth]{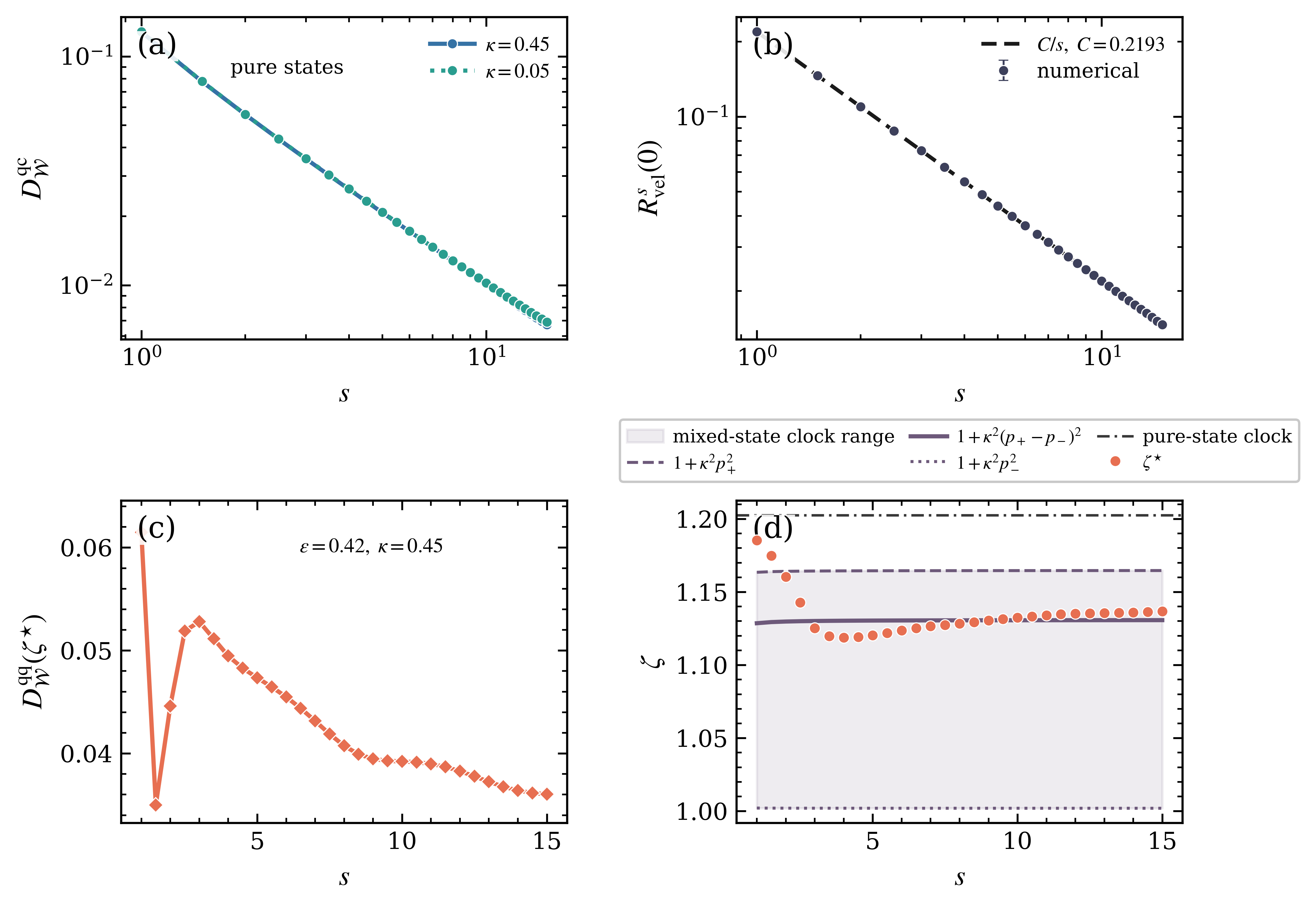}%
}
\caption{\textbf{Classicality measures with increasing spin.} 
(a) Mismatch between the post-transient quantum and classical dipoles, reached via $q$-LL and LL dynamics, respectively, as a function of $s$. Both dynamics are initialized as local pure states [$\varrho_j^2(0)=\varrho_j(0)$, corresponding to $\varepsilon=0$ in Eq.~(\ref{eq:initial-preparation})]. The mismatch is defined by Eq.~(\ref{eq:time-window-D}) and evaluated over time windows $\mathcal{W}=[0.3, 1]$~ns for $\kappa=0.45$, and $\mathcal{W}=[1, 2]$~ns for $\kappa=0.05$, both with fixed $\zeta=1$. 
(b) Initial dipole velocity mismatch $R^s_{\rm vel}(0)$ between the quantum and classical dynamics as a function of $s$, using the same parameters as in panel (a). The black dashed line represents the scaling $C/s$, with $C \approx 0.2193$ determined by the Hamiltonian parameters (see text). 
(c) Trajectory mismatch between the $q$-LL and $q$-LLG dynamics as a function of $s$. The system is initialized as a mixed state [$\varepsilon=0.42$ in Eq.~(\ref{eq:initial-preparation})] and evaluated over $\mathcal{W}=[0, 0.3]$~ns using pointwise-optimized $\zeta^{\star}$ values, with parameters $\varsigma_0=0.95$ and $\kappa=0.45$. 
(d) Pointwise-optimized $\zeta^{\star}$ values (fitted clock) for the cases shown in panel (c), compared against the clock factors $c_{+0}=1+\kappa^2p_{+}^2$, $c_{-0}=1+\kappa^2p_{-}^2$, and $c_{+-}=1+\kappa^2(p_+-p_-)^2$. The dashed-dotted line indicates the exact pure-state clock limit, $1+\kappa^2$.
The parameters entering the local mean-field Hamiltonian $h_j$ in Eq.~(\ref{eq:local-ham-bench2}) are $J=0.35$, $K_x=0.30$, $K_z=-0.12$, and $\mathbf B=\left(1,\frac{3}{20},1\right)$, with $\mathbf b=\mathbf B/|\mathbf B|$.}
\label{fig:benchmark2}
\end{figure*}

Building upon the same two-site model used in Benchmark 1, generalized to spin $s$, here we aim to test two further features of the product-state dynamics: the inequivalence between the $q$-LL and $q$-LLG constructions at finite spin $s$ (see Section \ref{sec:inequivalence}), and the quantum-classical correspondence as $s\to\infty$. 

For an arbitrary $s$, the local effective Hamiltonian, up to a scalar identity term, is
\begin{equation}
\label{eq:local-ham-bench2}
h_j=E_0\left[-(\mathbf b+J\mathbf n_k)\cdot\mathbf S_j
+\frac{K_x}{s}S_{x,j}^2+\frac{K_z}{s}S_{z,j}^2\right],
\quad k\neq j,
\end{equation}
which corresponds to an exchange interaction $-(E_0J/s)\mathbf{S}_1\cdot\mathbf{S}_2$ between the two quantum spins. The factors $1/s$ in Eq. (\ref{eq:local-ham-bench2}) keep the characteristic precession time of the normalized dipoles of order of $\hbar/E_0$ as $s$ increases. The classical reference system replaces $\mathbf{S}_j/s$ by vectors $\mathbf{m}_j$ with fixed magnitude in the corresponding spin Hamiltonian.

Most of these characterizations can be quantified by defining the notion of  \textit{distances} between the projected classical LL/LLG trajectories and/or their corresponding quantum trajectories:
\begin{equation}
\label{eq:distances}
D^{AB}(t,\zeta)=\left[\frac{1}{N}\sum_{j=1}^N\left|\boldsymbol{\mathfrak{m}}_j^{A}( t)-\boldsymbol{\mathfrak{m}}_j^{B}(\zeta t)\right|^2\right]^{1/2},
\end{equation}
which is a dynamical quantity. In Eq. (\ref{eq:distances}), $\{A,B\}$ classify if the comparison is between a quantum (q) or classical (c) simulation, and $\boldsymbol{\mathfrak{m}}_j(t)$ is the instantaneous normalized $j$-th dipole, which can be of the form of $\boldsymbol{\mathfrak{m}}_j^{\rm q}\equiv\mathbf{n}_j$ if quantum [see Eq. (\ref{eq:normalized-moment-quantum})], or $\boldsymbol{\mathfrak{m}}_j^{\rm c}\equiv\mathbf{m}_j$ if classical, and $\zeta>0$ rescales the time of trajectory $B$. From Eq. (\ref{eq:distances}), we can thus define a relevant quantity for our analysis. For a given observation window $\mathcal{W}=[
T_0,T]$, 
\begin{equation}
\label{eq:time-window-D}
D_{\mathcal{W}}^{AB}(\zeta,T_0,T)=\left[\frac{1}{T-T_0}\int_{T_0}^{T}D^{AB}(t,\zeta)^2dt\right]^{1/2},
\end{equation}
is the time-averaged trajectory distance. The quantum and classical dipole dynamics for each corresponding pair of equations of motion (local $q$-LLG/LLG or local $q$-LL/LL) are compared at the same physical time, with $\zeta=1$. In turn, to compare the dynamics generated by the two derived dissipative quantum laws, we instead optimize a single constant clock $\zeta^{\star}$ separately for each spin $s$,
\begin{equation}
\zeta^{\star}(s)=\underset{\zeta}{\arg\min}\left[
D_{\mathcal W}^{\rm qq}(\zeta)\right].
\end{equation}

We consider the following setup. Let $|\theta,\varphi\rangle_s\equiv\mathfrak{R}(\theta,\varphi)|s,s\rangle$ and the projector $P_s(\theta,\varphi)\equiv|\theta,\varphi\rangle_s\langle\theta,\varphi|_s$, with unit direction $s^{-1}{\rm Tr}[P_s(\theta,\varphi)\mathbf{S}]=\mathbf{u}(\theta,\varphi)=(\sin\theta\cos\varphi,\sin\theta\sin\varphi,\cos\theta)$. At each site $j=\{1,2\}$, the initial preparation is:
\begin{equation}
\label{eq:initial-preparation}
\varrho_j(0)=(1-\varepsilon)P_s(\theta_j,\varphi_j)+\varepsilon P_s(\theta_j+\varpi_j\varsigma_s,\varphi_j),
\end{equation}
where $\varpi_j=(-1)^{j-1}$, $\varsigma_s=\varsigma_0/\sqrt{s}$, and $\varsigma_0=0.95$. Thus, a pure state can be produced with $\varepsilon=0$ or $\varepsilon=1$, while $0<\varepsilon<1$ produces a mixed state characterized by an incoherent mixture of two spin-coherent states. This choice is based on the following properties: (\textit{i}) for any $0\leq \varepsilon\leq 1$, and by using that \cite{peres95}
\begin{eqnarray}
\left|{}_s\langle \theta,\varphi \ket{\theta',\varphi'}_s\right|^2 =  
\left[ \frac{1+\mathbf{u}(\theta,\varphi)\cdot \mathbf{u}(\theta',\varphi')}{2} \right]^{2s} , 
\end{eqnarray}
one obtains the purity 
\begin{equation}
\label{eq:purity-initial-state}
{\rm Tr}\varrho_j^2(0)={\rm Tr}\varrho_j^2(t)=1-2\varepsilon(1-\varepsilon)\left[1-\cos^{4s}\left(\frac{\varsigma_0}{2\sqrt{s}}\right)\right],
\end{equation}
which has the asymptotic value ${\rm Tr}\varrho_j^2=1-2\varepsilon(1-\varepsilon)(1-e^{-\varsigma_0^2/2})\leq1$ as $s\to\infty$; and (\textit{ii}) the magnitude of the initial normalized dipole 
\begin{eqnarray}
\mathbf{n}_j(0) & = & (1-\varepsilon)\mathbf{u}(\theta_j,\varphi_j) +
\varepsilon\mathbf{u} (\theta_j+\varpi_j\varsigma_s,\varphi_j) 
\end{eqnarray} 
for large $s$, is given by
\begin{equation}
\label{eq:initial-dipole}
|\mathbf{n}_j(0)| = 1-\frac{\varepsilon(1-\varepsilon)\varsigma_0^2}{2s} + \mathcal{O}\left(\frac{1}{s^2}\right),
\end{equation}
and therefore $|\mathbf{n}_j(0)|\rightarrow 1 $ as $s\to\infty$. 
This means that, in such preparation and in the high-spin limit, the local magnetization approaches saturation even though the purity does not approach unity when $0< \varepsilon <1$.

Panels (a) and (b) of  Fig.~\ref{fig:benchmark2} illustrate the pure-state ($\varepsilon = 0$) dynamics. Fig. \ref{fig:benchmark2}(a), compares the post-transient quantum and classical dipoles at the same physical time, with $\zeta=1$. For this coherent preparation, the mismatch obtained via the $q$-LL and classical LL dynamics scales as $s^{-1}$, and the two damping values ($\kappa=0.05$ and $\kappa=0.45$) give nearly overlapping curves. Thus, the quantum evolution progressively approaches the classical reference. This convergence is further corroborated by Fig.~\ref{fig:benchmark2}(b), which evaluates the initial velocity mismatch between the quantum and classical trajectories:
\begin{equation}
\label{eq:initial-velocity-mismatch-definition}
R^s_{\rm vel}(0)=\left[\frac{\sum_{j=1}^2 |\dot{\mathbf{n}}_j(0)-\dot{\mathbf{m}}_j(0)|^2}{\sum_{j=1}^2|\dot{\mathbf{m}}_j(0)|^2}\right]^{1/2}=\frac{C}{s},
\end{equation}
where the constant $C \approx 0.2193$ \footnote{The exact value of $C$ is:
\begin{equation}
\label{eq:constant-c}
C=\left[\frac{\sum_j|\mathbf{n}_j(0)\times \mathsf{K}\mathbf{n}_j(0)|^2}{\sum_j|\mathbf{n}_j(0)\times\mathbf{B}_j(0)|^2}\right]^{1/2},
\end{equation}
where $\mathsf{K}={\rm diag}(K_x,0,K_z)$ is the anisotropy matrix used, and $\mathbf{B}_j(0)$ is the initial effective field on spin $j$.} is determined exactly from the spin Hamiltonian parameters and the initial coherent-state orientations. Taken together, these results indicate the convergence of the coherent-state quantum dynamics to its classical counterpart, demonstrating asymptotic agreement in both the initial velocities and the final post-transient states.

In the mixed state ($0 < \varepsilon < 1$) dynamics, each local density matrix has exactly three distinct eigenvalues, being the complete spectrum given by $\{p_{+},p_{-},\underbrace{0,\ldots,0}_{d-2}\}$, where $p_{\pm}$ are functions of $s$:
\begin{equation}
\label{eq:nonzero-eigenvalues-mixed}
p_{\pm}(s)\equiv p_{\pm}^s=\frac{1}{2}\left\{1\pm\sqrt{1-4\varepsilon(1-\varepsilon)\left[1-\cos^{4s}\left(\frac{\varsigma_0}{2\sqrt{s}}\right)\right]}\right\}.
\end{equation}
In other words, the dynamics is characterized by three clock factors: $c_{+0}=1+\kappa^2p_{+}^2$, $c_{-0}=1+\kappa^2p_{-}^2$, and $c_{+-}=1+\kappa^2(p_{+}-p_{-})^2$ -- here independent of site $j$ since  $\kappa_j=\kappa$ is assumed uniform across both quantum spins. The first two connect occupied eigenspaces with the zero-eigenvalue subspace, while the third connects the two occupied eigenspaces; such as in Eq. (\ref{eq:purity-initial-state}), all nonzero eigenvalues from Eq. (\ref{eq:nonzero-eigenvalues-mixed}) and the three clock factors $c$ have an associated asymptotic value with increasing spin $s$: $p_{\pm}^{\infty}$, and $c_{+0}^{\infty}$, $c_{-0}^{\infty}$, $c_{+-}^{\infty}$. Since even in the large-$s$ limit a finite gap remains, the transformation between $q$-LL and $q$-LLG cannot be represented by a single scalar factor. 

Figure \ref{fig:benchmark2} (c) shows exactly that. For $\varepsilon=0.42$, the distance $D^{qq}_{\mathcal{W}}(\zeta^{\star},T)$, computed with the optimized time rescaling parameter $\zeta^{\star}$ and the observation window $\mathcal{W}=[0,T]$, with $T=0.3$ ns, is nonmonotonic at small $s$, and decreases slowly, keeping finite at $s=15$. In this case, although an observer would see the normalized dipole tending to saturation -- and thus asymptotically with a well-specified orientation on the unit sphere
--, locally the trajectories in state space would be different and strongly dependent on the dissipation structure chosen to model the system. Therefore, the saturation of the dipole itself is not sufficient to guarantee the collapse of the equations of motion in a single structure well-defined by a single clock. This interpretation is corroborated by Fig. \ref{fig:benchmark2} (d), in which we see that, for small spin $s$, the optimal $\zeta^{\star}$ parameter is closer to the pure-state reference scalar $\zeta^{\star}=1+\kappa^2=1.2025$ for $s=\frac{1}{2}$, and approaches values within the interval $c_{-0}^{\infty}\leq\zeta^{\star}\leq c_{+0}^{\infty}$ as $s$ increases, becoming particularly close to $c_{+-}^{\infty}$, consistent with an increasing role of $+\leftrightarrow-$ transition channel in the dipole dynamics. Despite optimization of $\zeta$, the distance $D^{qq}_{\mathcal{W}}(\zeta^{\star},T)$ over the entire simulation time remained finite \footnote{In fact, for trajectories starting from the same state, Taylor expansion gives
\begin{equation}
\begin{gathered}
\min_{\zeta}D_{[0,T]}^{qq}(\zeta) \\ =\frac{T}{\sqrt{3}}\min_\zeta\left[\frac{1}{N}\sum_{j=1}^N\left|\dot{\mathbf{n}}_j^{q-{\rm LL}}(0)-\zeta\dot{\mathbf{n}}_j^{q-{\rm LLG}}(0)\right|^2\right]^{1/2} \\ +\mathcal{O}(T^2),
\end{gathered}
\end{equation}
where the factor $1/\sqrt{3}$ comes from $T^{-1}\int_0^T t^2dt=T^2/3$.
}. 


\subsection{Benchmark 3: the large-$N$ limit}

\begin{figure*}[htb]
\centering
\makebox[\textwidth][c]{%
  \includegraphics[width=0.85\textwidth]{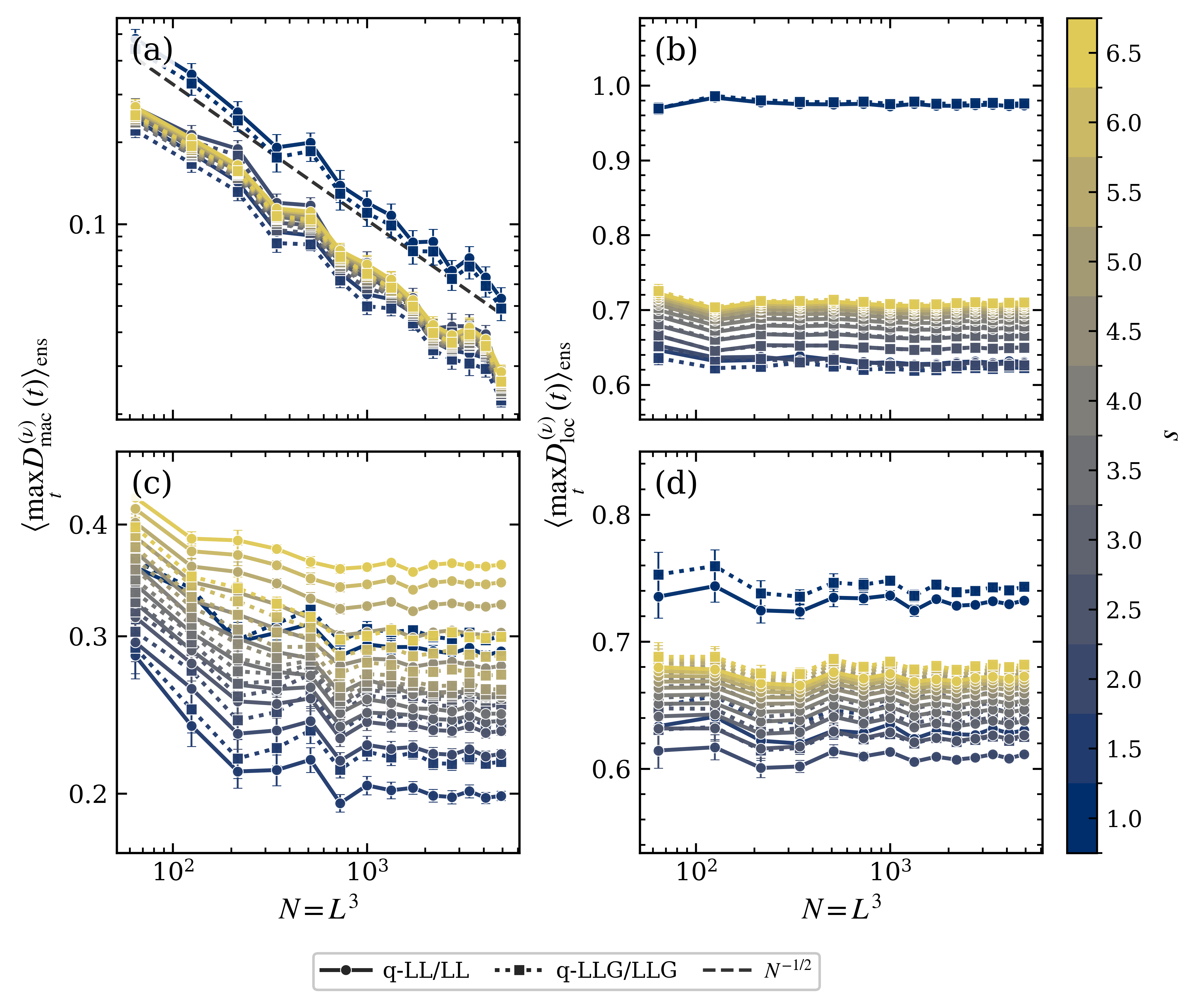}%
}
\caption{\textbf{Macroscopic and local quantum-classical mismatch with increasing system size.} (a) and (c) show the ensemble mean of the maximum macroscopic discrepancy between quantum and classical trajectories, $\langle\max_{t}D_{\rm mac}^{(\nu)}(t)\rangle_{\rm ens}$, for the first 0.3 ns of dynamics for initial mixed states as a function of the number of sites $N=L^3$, for a box of size $L$; (b) and (d) show the corresponding ensemble mean of the maximum local trajectory discrepancy, $\langle\max_{t}D_{\rm loc}^{(\nu)}(t)\rangle_{\rm ens}$, also as a function of $N$ for the corresponding systems in panels (a,c). In panels (a) and (b), the global $\pi$-rotation symmetry shared by the Hamiltonian and the initial ensemble enforces that the ensemble-averaged quantum and classical dipoles vanish: $\langle\mathbf{n}_j(t)\rangle_{\rm ens}=\langle\mathbf{m}_j(t)\rangle_{\rm ens}=0$. In panels (c) and (d), an applied magnetic field and a biased initial ensemble remove this constraint, allowing for nonzero ensemble-averaged dipoles. All trajectories start from matched local dipoles, $\mathbf{n}_j(0)=\mathbf{m}_j(0)$, $\forall j$. Colors indicate the spin $s$, solid lines with circles compare $q$-LL with LL, and dotted lines with squares compare $q$-LLG with LLG. The error bars indicate standard errors across 32 independent realizations for each spin $s$ and spin quantity $N$. The dashed $N^{-1/2}$ line in panel (a) is a guide for vanishing local discrepancy $|\boldsymbol{\mu}_N|^2$ but finite spatial correlation. All parameters except $N$ are kept fixed within each spin series. The largest system size evaluated comprised $N=17^3=4913$ quantum spins. The parameters entering the local mean-field Hamiltonian $h_j$ in Eq.~(\ref{eq:local-ham-benchmark3}) are $J=0.28$, $K_x=0.22$, and $K_z=-0.10$.}
\label{fig:benchmark3}
\end{figure*}

We next analyze the large $N$ limit. Here, we consider a simple cubic lattice with $N=L^3$ sites and periodic boundary conditions. We extend the two-site model of Benchmark 2 to include nearest-neighbor exchange interactions, and keep the same quadratic single-ion anisotropy and spin normalization:
\begin{equation}
\label{eq:hamiltonian-benchmark3}
\mathscr{H}=E_0\left\{\sum_j\left[-\mathbf{b}\cdot\mathbf{S}_j+\frac{K_x}{s}S_{x,j}^2+\frac{K_z}{s}S_{z,j}^2\right]-\frac{J}{zs}\sum_{\langle jk\rangle}\mathbf{S}_j\cdot\mathbf{S}_k\right\},
\end{equation}
where $z=6$ is the coordination number, and each nearest-neighbor pair $\langle jk\rangle$ is counted once. The normalization of $J$ by $z$ makes the exchange interaction scale associated with the mean neighboring dipole. Indeed, Eq. (\ref{eq:hamiltonian-benchmark3}) gives
\begin{equation}
\label{eq:local-ham-benchmark3}
h_j=E_0\left[-\left(\mathbf{b}+\frac{J}{z}\sum_{k\in\mathcal{N}(j)}\mathbf{n}_k\right)\cdot\mathbf{S}_j+\frac{K_x}{s}S_{x,j}^2+\frac{K_z}{s}S_{z,j}^2\right],
\end{equation}
where $\mathcal{N}(j)$ denotes the nearest neighborhood of site $j$, and $\mathbf{n}_k$ follows the definition of Eq. (\ref{eq:normalized-moment-quantum}). 

We initialize each local quantum spin in the maximum entropy state compatible with a prescribed polarization. This avoids introducing arbitrary higher multipole moments and ensures that the initial quantum and classical dipoles are exactly identical. Explicitly, the local quantum spins are initialized as
\begin{equation}
\label{eq:initial-state-benchmark-3}
\varrho_j^{(\nu)}(0)\equiv \frac{1}{Z_s}\exp\left(\lambda_s\frac{\mathbf{e}_j^{(\nu)}\cdot\mathbf{S}}{s}\right),
\end{equation}
where $\mathbf{e}_j^{(\nu)}$ is the initial unit direction of spin $j$ at realization $\nu$, $Z_s\equiv Z_s(\lambda_s)=\sum_{m=-s}^{s}\exp(\lambda_s m/s)$, and $\lambda_s$ is a Lagrange multiplier that fixes the polarization $r$ such that 
\begin{eqnarray}
r\mathbf{e}_j^{(\nu)} = \frac{1}{s}\mathrm{Tr}[\varrho_j^{(\nu)}(0)\mathbf{S}].  
\end{eqnarray}

The local purity is fixed and given by $\mathrm{Tr}\left([\varrho_j^{(\nu)}]^2\right)=Z_s(2\lambda_s)/Z_s^2(\lambda_s)$, which implies that the purity of the entire system vanishes in the thermodynamic limit: for each realization $\nu$, $\mathrm{Tr}(\varrho^2)=[\mathrm{Tr}(\varrho_j^{2})]^N\to 0$ as $N\to\infty$. We set $r=0.7$, sufficiently far from both the unpolarized limit ($r=0$), where the classical LL/LLG equations become singular, and the saturated limit ($r=1$), where the maximum entropy state reduces to a pure spin coherent state. The unit directions $\mathbf{e}_j^{(\nu)}=\mathbf{g}_j^{(\nu)}/|\mathbf{g}_j^{(\nu)}|$ are constructed from independent isotropic Gaussian vectors $\boldsymbol{\xi}_j^{(\nu)}$ via
\begin{equation}
\label{eq:gj}
\mathbf{g}_j^{(\nu)}=\boldsymbol{\xi}_j^{(\nu)}+w\sum_{k\in \mathcal{N}(j)}\boldsymbol{\xi}_k^{(\nu)}+\mathbf{a},
\end{equation}
where $\mathcal{N}(j)$ is the set of nearest neighbors of site $j$. We initially set the bias $\mathbf{a}=0$ and $w=0.35$; taking $w\neq 0$ introduce short-range orientational correlations, motivated by the short-range order that is naturally present in itinerant magnets and can persist even when long-range order is absent \cite{Antropov2005}. 
The lattice spacing, interaction parameters, initial polarization, and observation time $T$ remain fixed with increasing $N$.

To distinguish between local and macroscopic mismatches of quantum and classical dynamics, based on the notion of distances introduced by Eq. (\ref{eq:distances}), we define the following quantities for each realization:
\begin{eqnarray}
\label{eq:mac-loc-definitions}
D_{\rm mac}^{(\nu)}(t) & = &\left|\frac{1}{N}\sum_{j=1}^N[\mathbf{n}_j^{(\nu)}(t)-\mathbf{m}_j^{(\nu)}(t)]\right|
\nonumber \\
D_{\rm loc}^{(\nu)}(t) & = & \left[\frac{1}{N}\sum_{j=1}^N|\mathbf{n}_j^{(\nu)}(t)-\mathbf{m}_j^{(\nu)}(t)|^2\right]^{1/2},
\end{eqnarray}
where $D_{\rm mac}^{(\nu)}$ measures, at realization $\nu$, the discrepancy between spatially averaged quantum and classical dipoles, including possible cancellations among sites -- being therefore sensitive to the discrepancy visible at the macroscopic level -- while $D_{\rm loc}^{(\nu)}$ captures the full site-resolved mismatch.

The results for a total observation time of $T=0.3$~ns are shown in Fig.~\ref{fig:benchmark3}. For each spin $s$ and system size $N$, we first compute the maximum of each distance in Eq. (\ref{eq:mac-loc-definitions}) over the observation window individually for each of the 32 independent realizations. Fig.~\ref{fig:benchmark3} shows the mean of these maxima, with $\nu$ labeling the realizations; the error bars indicate the standard errors of the mean over all realizations. In panels (a) and (b), both the Hamiltonian and the initial ensemble share a $\pi$-rotation symmetry about the principal spin axes. These symmetries force both the quantum and classical ensemble-averaged dipoles to vanish at every site and time step for any finite $N$: $\langle\mathbf{n}_j(t)\rangle_{\mathrm{ens}}=\langle\mathbf{m}_j(t)\rangle_{\mathrm{ens}}=0$, where $\langle\cdot\rangle_{\mathrm{ens}}$ denotes the ensemble average.

However, this ensemble average does not itself eliminate the contribution of spatial correlations to $\langle \max_t D_{\rm mac}^{(\nu)}(t) \rangle_{\rm ens}$. To quantify this contribution, we can analyze the second moment, i.e., the ensemble mean of the squared distance at fixed time $t$, $\langle D_{\rm mac}^{2}(t)\rangle_{\rm ens}$ -- hereafter omitting the  label $(\nu)$ for brevity. Its square root gives an upper bound on the mean distance at that time, providing a basis for interpreting Fig. \ref{fig:benchmark3}. Defining the difference vector $\mathbf{d}_j(t)\equiv\mathbf{n}_j(t)-\mathbf{m}_j(t)$, we can write
\begin{equation}
\label{eq:d-mac-corr}
\langle D_{\mathrm{mac}}^2(t)\rangle_{\mathrm{ens}}=|\boldsymbol{\mu}_N(t)|^2+\frac{1}{N}\sum_{\mathbf{r}}\mathcal{C}_N(\mathbf{r},t),
\end{equation}
where $\boldsymbol{\mu}_N(t)\equiv\langle\mathbf{d}_j(t)\rangle_{\mathrm{ens}}$ (identical for all spins $j$), and
\begin{equation}
\label{eq:corr-def}
\mathcal{C}_N(\mathbf{r},t)=\frac{1}{N}\sum_j\langle\mathbf{d}_j(t)\cdot\mathbf{d}_{j+\mathbf{r}}(t)\rangle_{\mathrm{ens}}-|\boldsymbol{\mu}_N(t)|^2
\end{equation}
is the spatial correlation function on the periodic lattice, with $\mathbf{r}$ denoting the separation vector between sites. Since the system in panels (a) and (b) is invariant under $D_2$, the first term in Eq.~(\ref{eq:d-mac-corr}) vanishes identically ($|\boldsymbol{\mu}_N(t)|^2=0$). At the same time, the second term survives for finite $N$. In the large-$N$ limit, the mismatches carried by distinct correlation volumes add incoherently, leading to the asymptotic self-averaging behavior $\langle \max_{t}D_{\mathrm{mac}}^{(\nu)}\rangle_{\mathrm{ens}}\sim\mathcal{O}(N^{-1/2})$, as observed in Fig.~\ref{fig:benchmark3}(a). Conversely, the local distance satisfies $\langle D_{\mathrm{loc}}^2(t)\rangle_{\mathrm{ens}}=\mathcal{C}_N(\mathbf{0},t)+|\boldsymbol{\mu}_N(t)|^2=\mathcal{C}_N(\mathbf{0},t)$ by definition when $|\boldsymbol{\mu}_N|^2$ vanishes, and therefore remains limited [$\mathcal{O}(1)$] and nonzero. Taken together, the top panels demonstrate that an \textit{apparent} quantum-classical agreement in the dipole dynamics can emerge from spatial averaging in the thermodynamic limit, provided correlations are sufficiently short-ranged. This agreement can thus arise without any modifications to the local dissipation laws. Interestingly, this macroscopic masking persists for full-rank local mixed states, and is therefore not restricted to pure spin-coherent preparations; the macroscopic agreement, however, coexists with a finite local mismatch in the individual trajectories. This mechanism may have contributed to the well-known empirical success of the classical LLG equation in describing macroscopic magnetic observables.

Panels (c) and (d) illustrate the scenario where we break the $D_2$ invariance of the upper row by combining a nonzero external magnetic field with components along all three principal axes ($\mathbf{b}\neq 0$) in Eqs. (\ref{eq:hamiltonian-benchmark3}) and (\ref{eq:local-ham-benchmark3}), and a biased initial ensemble [i.e., $\mathbf{a}\neq 0$ in Eq.~(\ref{eq:gj})]. In this case, spatial averaging as $N$ increases also suppresses the correlation term in Eq.~(\ref{eq:d-mac-corr}), but the local discrepancy $|\boldsymbol{\mu}_N|^2$ remains finite. As a result, the masking effect is not only reduced but qualitatively distinct: in this case, the ensemble average $\langle \max_{t}D_{\mathrm{mac}}^{(\nu)}(t)\rangle_{\mathrm{ens}}$ tends to a nonzero limit with increasing $N$. By favoring particular spin orientations, bias sources -- or, more generally, $D_2$-invariance-breaking stimuli such as an applied magnetic field -- prevent local quantum-classical discrepancies from canceling completely in the spatial average, producing distinct macroscopic dynamics that persist as the system size increases.

This complements Benchmark 2, where the local spin dimension and initial preparation are varied. Increasing $s$ can improve selected coherent-state comparisons, while increasing $N$, as demonstrated in Benchmark 3, can improve the quantum-classical agreement because local differences cancel out when spatially averaged, even though the trajectories at individual sites remain distinct.


\section{Conclusions}
\label{sec:conclusions}
We have developed a scalable variational framework for quantum atomistic spin dynamics based on a product-state \textit{ansatz} that retains the full local density operator while neglecting intersite entanglement, for both pure and mixed local states. This separates local quantum dynamics from many-body effects driven by entanglement and provides an effective quantum extension of classical spin dynamics. The resulting conservative part of the dynamics follows from the Dirac-Frenkel variational principle, while dissipation can be introduced through either a Rayleigh-Lagrange-d'Alembert or an Onsager construction, yielding the local $q$-LLG and $q$-LL dynamics, respectively. These equations share the same conservative limit but describe generally different dissipative dynamics.

The inequivalence of $q$-LL and $q$-LLG has a geometric origin. For a qubit, and more generally for a local state with exactly two distinct eigenvalues, the two local dynamics become equivalent up to a rescaling of time. For higher dimensional mixed states, distinct spectral gaps lead to different relaxation of tangent directions, preventing such a universal rescaling. Thus, the quantum LL-LLG distinction is intrinsically tied to the structure of the local state space.

The results identify two distinct routes to quantum-classical agreement. In the large $s$ limit, quantum corrections to the dipole dynamics are suppressed for coherent local states, with both the initial velocity and post-transient dipole mismatches in the considered model scaling as $1/s$. This constitutes a genuine local classical limit. For mixed initial preparations, however, a nonzero $q$-LL/$q$-LLG trajectory mismatch remains even after clock optimization. By contrast, increasing the system size can produce classical behavior at the macroscopic level through self-averaging. For a vanishing ensemble-average difference between quantum and classical trajectories and sufficiently short-ranged correlations, the discrepancy between quantum and classical dipole dynamics shows a $N^{-1/2}$ scaling for spatially averaged observables, even though the local discrepancy remains $\mathcal{O}(1)$.

These limits reflect two fundamentally different mechanisms: large $s$ suppresses local quantum corrections in coherent initial preparations, whereas large $N$ suppresses their contribution to macroscopic observables through collective averaging. As a consequence, classical macroscopic behavior does not necessarily imply a classical microscopic dynamics, and a finite local quantum correction does not necessarily produce a measurable macroscopic discrepancy. This distinction provides a possible explanation for why classical atomistic spin models proved so successful in reproducing spatially averaged magnetic observables even when local quantum effects remain significant.

\section*{Acknowledgements}
I.P.M. acknowledges financial support from the Crafoord Foundation through grant No. 20231063. E.S. acknowledges financial support from the Swedish Research Council (VR) 
through Grant No. 2025-05249. The computational resources were provided by the National Academic Infrastructure for Supercomputing in Sweden (NAISS), funded by the Swedish Research Council. 

\section*{Appendix}
\appendix

\section{Action on manifold $\mathcal{M}$ via purification}
\label{sec:purification-scheme}

Let $\varrho_j(0)=\sum_{a=1}^d p_{j,a}|a_j\rangle\langle a_j|$ be the spectral decomposition of $\varrho_j(0)$ for a set of orthonormal states $|a_j\rangle$ in the local Hilbert space $\mathcal{H}_j$, with coefficients $p_{j,a}\geq0$ such that $\sum_{a}p_{j,a}=1$. Let also $\tilde{\mathcal{H}}_j$ be a $d$-dimensional auxiliary Hilbert space associated with site $j$, and define the local purification in $\mathcal{H}_j\otimes\tilde{\mathcal{H}}_j$ as
\begin{equation}
\label{eq:purified-initial-state}
|\Omega_j(0)\rangle=\sum_{a=1}^d\sqrt{p_{j,a}}|a_j\rangle\otimes|\tilde{a}_j\rangle,
\end{equation}
where $\{|\tilde{a}_j\rangle\}$ denotes an orthonormal basis of $\tilde{\mathcal{H}}_j$, such that ${\rm Tr}_{\tilde{\mathcal{H}}_j}(|\Omega_j(0)\rangle\langle\Omega_j(0)|)=\varrho_j(0)$. To represent the evolution of the local density operator in its isospectral orbit, a corresponding purified trajectory may be chosen as $|\Omega_j(t)\rangle=(\mathcal{U}_j\otimes\mathbbm{1}_{\tilde{\mathcal{H}}_j})|\Omega_j(0)\rangle$, resulting in
\begin{equation}
{\rm Tr}_{\tilde{\mathcal{H}}_j}(|\Omega_j(t)\rangle\langle\Omega_j(t)|)=\mathcal{U}_j{\rm Tr}_{\tilde{\mathcal{H}}_j}(|\Omega_j(0)\rangle\langle\Omega_j(0)|)\mathcal{U}_j^{\dagger}=\varrho_j(t).
\end{equation}
The purification given by Eq. (\ref{eq:purified-initial-state}) is not unique: an additional unitary transformation $\mathcal{V}_j$ acting solely on $\tilde{\mathcal{H}}_j$ leaves the local density operator unchanged. We fix this auxiliary unitary freedom by choosing $\mathcal{V}_j(t)\equiv\mathbbm{1}_{\tilde{\mathcal{H}}_j}$, as above \footnote{To make the auxiliary unitary freedom explicit, consider $|\Omega_j(t)\rangle=(\mathcal{U}_j\otimes\mathcal{V}_j)|\Omega_j(0)\rangle$, where $\mathcal{V}_j(t)$ is a differentiable unitary operator on $\tilde{\mathcal{H}}_j$, with $\mathcal{V}_j(t=0)=\mathbbm{1}_{\tilde{\mathcal{H}}_j}$. Unitarity gives
\begin{equation}
{\rm Tr}_{\tilde{\mathcal{H}}_j} \ket{\Omega_j} \bra{\Omega_j} = \mathcal{U}_j \varrho_j(0) \mathcal{U}_j^{\dagger},
\end{equation}
so the local state $\varrho_j(t)$ is independent of $\mathcal{V}_j$. As a consequence, the energy $E[\{\varrho_k(t)\}]$ is also independent of the auxiliary unitaries, since the Hamiltonian acts only on the non-auxiliary (physical) degrees of freedom. However, in this case, the kinematic term separates as
\begin{equation}
\label{eq:kinematic-term-with-vj}
\begin{gathered}
i\hbar\langle\Omega_j\ket{\dot{\Omega}_j} = i\hbar\operatorname{Tr}_{j}\left( \varrho_j(0) \mathcal{U}_j^\dagger \dot{\mathcal{U}}_j\right) \\
+i\hbar \operatorname{Tr}_{\tilde{\mathcal{H}}_j}
\left(\tilde{\varrho}_j(0) \mathcal{V}_j^\dagger\dot{\mathcal{V}}_j
\right),
\end{gathered}
\end{equation}
where $\tilde{\varrho}_j(0)={\rm Tr}_{\mathcal{H}_j}|\Omega_j(0)\rangle\langle\Omega_j(0)|$. If $\mathcal{V}_j(t)$ is assumed independent of $\{\mathcal{U}_k(t)\}$, the second term of Eq. (\ref{eq:kinematic-term-with-vj}) contributes nothing to the variation of the action with respect to any $\mathcal{U}_k$. Therefore, the equations of motion for the local states $\varrho_j(t)$ remain unchanged. Furthermore, $\mathcal{U}_j\otimes\mathcal{V}_j$ preserves the Schmidt coefficients $\sqrt{p_{j,a}}$ of the local purification $|\Omega_j(0)\rangle$:
\begin{equation}
|\Omega_j(t)\rangle=\sum_{a=1}^{d}\sqrt{p_{j,a}}\left(\mathcal{U}_j|a_j\rangle\right)\otimes\left(\mathcal{V}_j|\tilde{a}_j\rangle\right),
\end{equation}
and therefore $\mathcal{V}_j$ does not alter the entanglement between the site $j$ and its auxiliary system. Also, because each $\mathcal{U}_j\otimes\mathcal{V}_j$ term acts only on $\mathcal{H}_j\otimes\tilde{\mathcal{H}}_j$, applying these transformations to an initial product of local purifications $|\Omega(0)\rangle=\bigotimes_j|\Omega_j(0)\rangle$ preserves the product form $\bigotimes_j|\Omega_j(t)\rangle$, and, after tracing out all auxiliary degrees of freedom, the original product state \textit{ansatz} $\bigotimes_j\varrho_j(t)$. Thus, choosing $\mathcal{V}_j\equiv \mathbbm{1}_{\tilde{H}_j}$ is a simple way of eliminating the auxiliary contribution to the kinematic term in Eq. (\ref{eq:kinematic-term-with-vj}) -- which, under the broad assumption that $\mathcal{V}_j$ is not a function of $\{\mathcal{U}_k(t)\}$, does not change any relevant result}.

In this case, taking for the complete system $|\Omega(t)\rangle = \bigotimes_{j}|\Omega_j(t)\rangle$, and tracing all auxiliary degrees of freedom in space $\tilde{\mathcal{H}}\equiv\bigotimes_j\tilde{\mathcal{H}}_j$, yields the following:
\begin{equation}
{\rm Tr}_{\tilde{\mathcal{H}}}|\Omega(t)\rangle\langle\Omega(t)|=\bigotimes_j\mathcal{U}_j(t)\varrho_j(0)\mathcal{U}_j^{\dagger}(t)=\bigotimes_j\varrho_j(t),
\end{equation}
while, for fixed coefficients $p_{j,a}$, and applying the action in Eq. (\ref{eq:action}) to $|\Omega(t)\rangle$, with the extended Hamiltonian $\mathscr{H}\otimes\mathbbm{1}_{\tilde{\mathcal{H}}}$, gives
\begin{eqnarray}
i\hbar \langle\Omega \ket{\dot{\Omega}} & = & \sum_j i\hbar{\rm Tr}_j\left(\varrho_j(0)\mathcal{U}_j^{\dagger}\dot{\mathcal{U}}_j\right),
\nonumber \\
\bra{\Omega} \mathscr{H} \otimes \mathbbm{1}_{\tilde{\mathcal{H}}} \ket{\Omega} & = & {\rm Tr}\left[\left(\bigotimes_j\varrho_j\right)\mathscr{H}\right]\equiv E[\{\varrho_j\}],
\end{eqnarray}
yielding 
\begin{equation}
\label{eq:action-2}
\mathcal{S}[\{\mathcal{U}_j\}]=\int dt\left\{\sum_j i\hbar{\rm Tr}_{j}\left[\varrho_j(0)\mathcal{U}_j^{\dagger}(t)\dot{\mathcal{U}}_j(t)\right]-E[\{\varrho_j\}]\right\}.
\end{equation}
Fixing the Schmidt coefficients $\sqrt{p_{j,a}}$ of the local purifications is, thus, \textit{exactly equivalent} to restricting the density operator $\varrho$ to the manifold $\mathcal{M}$. 

Note that Eq.~\eqref{eq:action-2} provides the natural expression for the action on $\mathcal{M}$, given the isospectral orbits defined in Eq. (\ref{eq:isospectral-orbit}) of the main text.

\section{First-order anti-coherence generation: a simple solvable example}
\label{sec:appendix-anticoherence}

A single spin $s=1$ provides an exactly solvable example of dynamics that preserves the purity of the (local) state while driving the system towards a nonmagnetic (vanishing) dipole ground state. As we shall see, this is a minimal model for the asymptotic quantum spin dynamics selection of a first-order anti-coherent spin state, with vanishing spin expectation value \cite{Baguette2017}, starting from a coherent state.

Consider the following spin Hamiltonian:
\begin{equation}
\mathscr{H}=KS_z^2,
\end{equation}
where $K>0$ has units of energy; setting $K\equiv E_0$ recovers the single-spin Hamiltonian in Eq. (\ref{eq:spin-1-ham-bench1}). This is the single-ion and uniaxial anisotropy term of the model considered in Eq.~\eqref{eq:spin-hamiltonian-benchmark1} of the main text, with the remaining interactions switched off. We initialize the spin in the coherent state polarized along the $+x$ direction:
\begin{equation}
\label{eq:initial-psi-appendix-b}
|\psi(0)\rangle=e^{-i\frac{\pi}{2}S_y}|1\rangle=\frac{1}{2}\left(|1\rangle+\sqrt{2}|0\rangle+|-1\rangle\right).
\end{equation}

Therefore, from $\varrho(0)=|\psi(0)\rangle\langle\psi(0)|$, the initial dipole is $\mathbf{n}(0)=\mathbf{e}_x$ and ${\rm Tr}\varrho^2(0)=1$. For a pure state, the $q$-LL equation with $\tilde{\lambda}=\kappa/\hbar$ is generated by \cite{liu24}:
\begin{equation}
\frac{\partial}{\partial t}|\psi\rangle=-\frac{i}{\hbar}\mathscr{H}|\psi\rangle-\frac{\kappa}{\hbar}\left(\mathscr{H}-\langle\mathscr{H}\rangle\right)|\psi\rangle,
\end{equation}
which has, for a time-independent Hamiltonian, the normalized solution:
\begin{equation}
\label{eq:solution-appendix-b}
|\psi(\tau)\rangle=\frac{e^{-(i+\kappa)\tau}(|1\rangle+|-1\rangle)+\sqrt{2}|0\rangle}{\sqrt{2(1+e^{-2\kappa\tau})}},
\end{equation}
where $\tau\equiv Kt/\hbar$ is the dimensionless time parameter. This produces the following dipole and energy as a function of $\tau$:
\begin{equation}
\label{eq:dipole-energy-appendix-b}
\mathbf{n}(\tau)=\frac{\cos(\tau)}{\cosh(\kappa\tau)}\mathbf{e}_x,\quad E(\tau)=\frac{K}{1+e^{2\kappa\tau}},
\end{equation}
while the purity ${\rm Tr}\varrho^2=1$ is conserved. Because the density operator is pure at all times, the equivalent $q$-LLG solutions can be obtained by replacing $\tau$ with $\tau/(1+\kappa^2)$ in Eqs. (\ref{eq:solution-appendix-b}) and (\ref{eq:dipole-energy-appendix-b}). Thus, for $\tau\to\infty$, the damped quantum equations of motion select the first-order anti-coherent state $|0\rangle$, with $\mathbf{n}=\langle\mathbf{S}\rangle\to0$ as $t\to\infty$.

\section{Numerical methods}
\label{sec:numerical-methods}

Benchmark 1 uses the adaptive eighth-order integrator DOP853 as implemented in \texttt{SciPy}~\cite{Virtanen2020}, with relative and absolute tolerances of $10^{-10}$ and $10^{-12}$, respectively. The coupled local density matrices are evolved simultaneously. The output spacings for the isolated quantum spin calculation [Fig.~\ref{fig:benchmark1}(a)] and for the two-spin calculations [Fig.~\ref{fig:benchmark1}(b-d)] are $10^{-13}$~s and $2.5\times10^{-13}$~s, respectively; these time intervals define the resolution of the stored trajectories, whereas the internal integration step is selected adaptively by the solver. The exact two-spin reference in Fig.~\ref{fig:benchmark1}(b), represented by the black dashed line, is obtained by diagonalizing the full Hamiltonian $\mathscr{H}$ and evolving the density matrix as $\varrho(t)=e^{-i\mathscr{H}t/\hbar}\varrho(0)e^{i\mathscr{H}t/\hbar}$. In this case, the local states $\varrho_1(t)$ and $\varrho_2(t)$ are obtained by tracing out the other spin: $\varrho_1(t)={\rm Tr}_2\varrho(t)$ and $\varrho_2(t)={\rm Tr}_1\varrho(t)$.

In turn, Benchmarks 2 and 3 employ an explicit, fixed-step, second-order midpoint scheme. Writing the local equations as $\dot{\varrho}_j=(i/\hbar)[\varrho_j,A_j]$, where $A_j$ is a Hermitian generator, the update from time step $t_n$ to $t_{n+1}$ is
\begin{eqnarray}
\label{eq:update-benchmarks23}
\varrho_j^{n+1/2} & = & e^{-iA_j^n\Delta t/(2\hbar)}\varrho_j^{n}e^{iA_j^n\Delta t/(2\hbar)},
\nonumber \\
\varrho_j^{n+1} & = & e^{-iA_j^{n+1/2}\Delta t/\hbar}\varrho_j^ne^{iA_j^{n+1/2}\Delta t/\hbar},
\end{eqnarray}
where $\Delta t\equiv t_{n+1}-t_n$ is the step size, here set between $5\times10^{-16}$~s and $10^{-14}$~s, depending on the simulation. Given the local states $\{\varrho_j\}$, the generator $A_j$ in Eq.~(\ref{eq:update-benchmarks23}) can be obtained for the local $q$-LL and $q$-LLG equations as follows:
\begin{eqnarray}
\label{eq:Lj-both}
A_j^{q-{\rm LL}}=h_j+i\kappa[\varrho_j, h_j],
\nonumber \\
A_j^{q-{\rm LLG}}=W_j\tilde{A}_j^{q-{\rm LLG}}W_j^\dagger,
\end{eqnarray}
where $h_j$ is computed using Eq.~(\ref{eq:delta-j-E-rearranged}) of the main text; $W_j$ is the unitary matrix that diagonalizes $\varrho_j$, i.e., $\varrho_j=W_jP_jW_j^{\dagger}$ with $P_j\equiv{\rm diag}(p_{j,1},\ldots,p_{j,d})$; and $\tilde{A}_j^{q-{\rm LLG}}$ is the matrix with elements
\begin{equation}
\left(\tilde{A}_j^{q-{\rm LLG}}\right)_{ab}=\frac{\left(W_j^{\dagger}h_jW_j\right)_{ab}}{1-i\kappa(p_{j,a}-p_{j,b})},
\end{equation}
where components satisfying $|p_{j,a}-p_{j,b}|\leq 10^{-12}$ are set to zero. Since $h_j$ depends on the current set of local states $\{\varrho_j\}$, obtaining $A_j$ at the midpoint in Eq.~(\ref{eq:update-benchmarks23}) -- i.e., the generator $A_j^{n+1/2}$ at time $t_n+\Delta t/2$ -- requires first calculating the quantum dipoles $\mathbf{n}_j^{n+1/2}\equiv s^{-1}{\rm Tr}(\varrho_j^{n+1/2}\mathbf{S})$ from the density matrix $\varrho_j^{n+1/2}$ evaluated in the first line of Eq.~(\ref{eq:update-benchmarks23}). The midpoint local states $\varrho_j^{n+1/2}$ are also necessary to evaluate the effective local mean-field Hamiltonian $h_j^{n+1/2}$ from Eq.~(\ref{eq:delta-j-E-rearranged}). This Hamiltonian is then used to finally obtain $A_j^{n+1/2}$ via Eq.~(\ref{eq:Lj-both}) for both quantum equations of motion. Matrix exponentials are evaluated by diagonalizing the Hermitian generators $A_j$. After each half and full update as defined in Eq.~(\ref{eq:update-benchmarks23}), the local density matrices are symmetrized as $\varrho_j\leftarrow(\varrho_j+\varrho_j^\dagger)/2$ and normalized by their computed traces to correct for spurious floating-point deviations from Hermiticity and unit trace.

For the classical trajectories, we use the corresponding rotation midpoint predictor and set $\alpha=\kappa$. Given the classical angular velocity $\boldsymbol{\omega}_j\equiv\mathbf{m}_j\times\dot{\mathbf{m}}_j/|\mathbf{m}_j|^2$, the equation of motion is $\dot{\mathbf{m}}_j=\boldsymbol{\omega}_j\times\mathbf{m}_j$. Denoting a Rodrigues rotation by an angle $|\boldsymbol{\omega}|\iota$ as $\mathscr{R}(\boldsymbol{\omega},\iota)$, where $\iota$ is a parameter in units of time, we first produce a half-step configuration at time $t_n+\Delta t/2$ via
\begin{equation}
\mathbf{m}_j^{n+1/2}=\mathscr{R}\left(\boldsymbol{\omega}_j^{n},\frac{\Delta t}{2}\right)\mathbf{m}_j^n.
\end{equation}
The angular velocities at the half-step are then updated as $\boldsymbol{\omega}_j^{n+1/2}=\mathbf{m}_j^{n+1/2}\times\dot{\mathbf{m}}_j^{n+1/2}/\left|\mathbf{m}_j^{n+1/2}\right|^2$, and used to rotate the original classical dipole vectors as
\begin{equation}
\mathbf{m}_j^{n+1}=\mathscr{R}\left(\boldsymbol{\omega}_j^{n+1/2},\Delta t\right)\mathbf{m}_j^n.
\end{equation}
This naturally preserves the prescribed magnitudes of the classical dipoles up to a floating-point error without any subsequent normalization.

For time-rescaled comparisons, such as in Figs. \ref{fig:benchmark2}(c,d), a single rescaling factor $\zeta^{\star}$ is fitted for each spin $s$ by minimizing the average distance defined in Eq. (\ref{eq:time-window-D}) over a given interval $[\zeta_{\rm min},\zeta_{\rm max}]$.

\setcounter{equation}{0}
\renewcommand{\theequation}{N\arabic{equation}}
\renewcommand{\theHequation}{notes.\arabic{equation}}

\bibliographystyle{apsrev4-2}
\bibliography{bib}
\end{document}